\documentclass[12pt]{article}
\usepackage{amsmath}
\usepackage{graphicx}
\usepackage{enumerate}
\usepackage{amsfonts}
\usepackage{natbib} 
\usepackage{url} 
\usepackage{bm}
\usepackage{booktabs}
\usepackage{colortbl}
\usepackage{appendix}
\usepackage{multirow}
\usepackage{array}
\usepackage{subcaption}
\usepackage[dvipsnames]{xcolor}
\RequirePackage[hidelinks,colorlinks=true,urlcolor=blue,citecolor=blue,urlcolor=blue]{hyperref}
\usepackage[total={170mm,230mm},
 left=25mm,
 top=20mm]{geometry}
\newcommand{\blind}{0}
\usepackage{xr}

\newcommand\given[1][]{\:#1\vert\:}

\definecolor{bluishdarkgreen}{RGB}{0,77,100}
\definecolor{darkgreen}{RGB}{0,100,0}
\definecolor{rev1}{RGB}{216,27,96}
\definecolor{rev2}{RGB}{30,136,229}
\definecolor{rev3}{RGB}{181,103,33}

\usepackage[acronym]{glossaries}
\newacronym{nwp}{NWP}{Numerical Weather Prediction}
\newacronym{wrf}{WRF}{Weather Research and Forecasting}
\newacronym{spen}{SPEN}{Scottish Power Energy Networks}
\newacronym{ecmwf}{ECMWF}{European Centre for Medium Range Weather Forecasting}
\newacronym{bma}{BMA}{Bayesian Model Averaging}
\newacronym{eps}{EPS}{Ensemble Prediction System}
\newacronym{ws10max}{$\text{ws}10_{\max}$}{maximum 10-meter wind speed}

\begin{document}

\def\spacingset#1{\renewcommand{\baselinestretch}%
{#1}\small\normalsize} \spacingset{1}









{
  \title{\bf
  Probabilistic forecasting of weather-driven faults in electricity networks: a flexible approach for extreme and non-extreme events
  }

  \author{
    Mateus Maia\thanks{Corresponding author. Email: mateus.maiamarques@glasgow.ac.uk}, Daniela Castro-Camilo, Jethro Browell\\
    School of Mathematics and Statistics, University of Glasgow
  }

  \maketitle
}

\if1\blind
{
  \begin{center}
    {\LARGE\bf Title}
\end{center}
} \fi

\begin{abstract}
Electricity networks are vulnerable to weather damage, with severe events often leading to faults and power outages. Timely forecasts of fault occurrences, ranging from nowcasts to several days ahead, can enhance preparedness, support faster response, and reduce outage durations. To be operationally useful, such forecasts must quantify uncertainty, enabling risk-informed resource allocation. We present a novel probabilistic framework for forecasting fault counts that captures typical and extreme events. Non-extreme faults are modeled linearly interpolating estimates from multiple additive quantile regressions, while extreme events are described through a discrete generalized Pareto distribution. To incorporate the impact of weather fluctuations, we use ensemble numerical weather predictions, which helps to quantify uncertainty in the forecasts. This approach is designed to provide reliable fault predictions up to four days ahead.
We evaluate the model through numerical experiments and apply it to historical fault data from two electricity distribution networks in Great Britain. The resulting forecasts demonstrate substantial improvements over business-as-usual and alternative modeling approaches. A practitioner trial conducted with Scottish Power Energy Networks from October 2024 to March 2025 further demonstrates the operational value of the forecasts. Engineers found them sufficiently reliable to inform decision-making, offering benefits to both network operators and electricity consumers.

\end{abstract}

\noindent%
{\it Keywords:} electricity networks, extreme value theory for counts, generalized additive models, probabilistic forecasting, quantile regression, weather-driven faults.
\vfill

\newpage
\spacingset{1.2} 
\section{Introduction}
\label{sec:intro}
Electricity is essential to modern life, and disruptions to supply can lead to significant human and economic impacts. As such, ensuring the resilience of electricity networks is a critical priority for network operators, regulators, and consumers alike. Severe weather events pose a particular threat to exposed infrastructure, including overhead lines and their supporting poles and pylons. A key element of resilience is preparedness. While it is not possible to prevent weather-related damage entirely, anticipating extreme conditions allows for precautionary measures that can enhance system robustness and support timely repairs. Recent work in the UK, such as \cite{cha2025assessing}, has evaluated Storm Éowyn to highlight cross-sectoral impacts on critical infrastructure, reinforcing the importance of anticipating disruptions and enhancing system resilience. This underscores the importance of forecasting the number and location of faults likely to occur in an electricity network over lead times ranging from hours to days, thereby enabling proactive planning and response.

While there is extensive literature on the resilience of electricity networks to adverse weather, most of it centres on asset design and long-term network planning. In contrast, research on short-term (days-ahead) fault prediction remains limited, and few commercial solutions currently exist in this area—offered by companies such as IBM, GE, DTN, Bellrock, and StormGeo. Existing methods are often limited to forecasting one day ahead only and offer no or limited uncertainty quantification. They do, however, provide strong evidence that weather-driven fault prediction is possible, and these methodologies are likely extendable to both longer-range and probabilistic forecasting.

The most comprehensively reported fault prediction capability is the University of Connecticut’s ``Outage Prediction Model''. Used by industry, this model predicts faults in the northeastern USA on a 4km grid during storms \citep{watson2020weather,cerrai2019storm}. It predicts outages---defined as any incident requiring a repair crew---exclusively for extra-tropical storms. Studies indicate that the errors in the weather forecast contribute only marginally to overall fault prediction errors, with the primary source of error stemming from the weather-to-fault modeling process. An extension incorporating thunderstorms was introduced in \cite{alpay2009thunder}. Faults on transmission towers caused by tropical cyclones in China are forecast in \cite{huang2025modeling} using fragility curves and down-scaled weather forecasts to address systematic bias in forecasts of extreme wind speeds.

Others have developed similar, albeit less sophisticated, models for other regions, such as Finland \citep{brester2020weather}. In the United Kingdom (UK), \cite{Tsioumpri2021WeatherNetworks} and \cite{Wilkinson2022ConsequenceStorms} developed a methodology that was later expanded into a commercial offering by DTN. The latter established a correlation between weather forecasts\textemdash{}specifically precipitation, wind speed and lightning\textemdash{} and the number of faults, proposing simple statistical models for predicting the number of faults in UK Power Networks’ distribution network in the southeast UK. Additional example include \cite{ghaemi2024stacking}, which performed fault forecasting based on weather predictors for power transmission line faults;
power outages in the Great Britain have also been associated with large-scale weather patterns which can be forecast with some skill up to a few weeks ahead \citep{souto2024identification}.

This work is motivated by the distribution of faults across the 11 districts within \gls{spen}. Under normal operational conditions, the control center can benefit from leveraging probabilistic forecasting tools to allocate resources more efficiently and schedule maintenance proactively, considering both typical and rare weather scenarios. Events characterized by a higher number of faults, classified as red events in a traffic light-style system, are typically associated with severe weather. These events are of particular interest due to the high preparedness required by the maintenance team, although they occur with relatively low frequency. On the other hand, forecasts for less disruptive but more frequent amber events, which involve a lower number of faults, are still important for daily management of resources.

\begin{figure}
    \centering    
    \includegraphics[width=\linewidth]{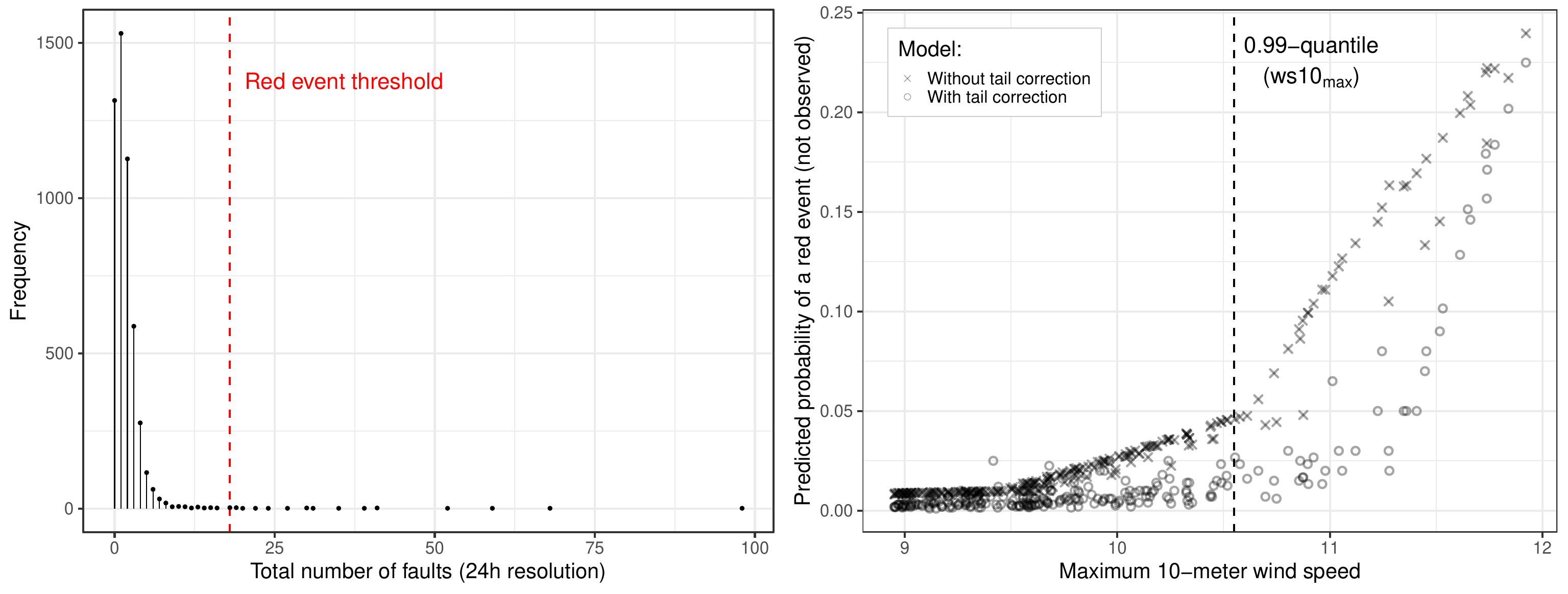}
    \caption{ Daily occurrence of faults in a SPEN district over ten years (left), with the vertical dashed line indicating the red--event threshold. The right panel shows estimated probabilities of issuing a red warning on non-event days when \gls{ws10max} exceeds 9m/s, comparing the naive and proposed models. }
    \label{fig:freq_faults_plot}
\end{figure}

The left panel of Figure~\ref{fig:freq_faults_plot} displays the daily occurrence of faults over a ten-year period in a single SPEN district, illustrating both typical and extreme fault scenarios. {The right panel compares the two modeling approaches by showing, for days with \gls{ws10max} exceeding approximately 9\,m/s, the predicted probability of issuing a red fault band warning on occasions where a red event did \emph{not} occur. The naive model—represented by crosses and not explicitly designed to account for extremes—tends to assign comparatively higher probabilities to unobserved red events. In contrast, the proposed approach in this manuscript—shown with circles—yields lower and more stable probability estimates in this regime, consistent with its mixture-based treatment of extreme and non-extreme behavior.}


In probabilistic forecasting problems, quantile regression (QR), initially developed by \cite{koenker1978regression}, has been widely applied including the energy sector \citep{zhu2024large,ruiz2024applications,zhang2014review}. The success of QR can be attributed to several factors: it effectively describes the distribution of new observations beyond just the mean, demonstrates robustness to outliers, and requires weaker distributional assumptions compared to traditional distributional regression models \citep{koenker2013distributional}. Moreover, recent advancements in QR, such as additive QR, have provided greater flexibility by incorporating additive terms, allowing for the specification of non-linear effects between predictors and estimated quantiles \mbox{\citep{fasiolo2021fast,waldmann2013bayesian}}.

Although QR is classically formulated for continuous response variables, it can be applied to count data when the primary focus is predictive performance than inference. This setting is relevant for fault forecasting in electricity networks, where accurate estimation of the quantiles is more critical than strict inferential analysis. For discrete responses, the presence of mass points and non-differentiability of the quantile function pose challenges to standard asymptotic theory for sample quantiles and associated inference \citep{machado2005quantiles}. To address this, approaches based on response jittering have been proposed, in which bounded continuous noise is added to the response to approximate a continuous distribution \citep{machado2005quantiles,jantre2023bayesian}. However, by introducing additional randomness, jittering can lead to biased quantile estimates and increased computational cost. When reliable prediction is the primary goal and formal inference is secondary, directly applying QR to count outcomes remains a pragmatic modeling choice.

Figure~\ref{fig:freq_faults_plot} shows that modeling fault counts requires explicit treatment of tail behavior. While quantile regression yields probabilistic forecasts, estimation of conditional extremes is limited by data sparsity in the tails, especially under covariate conditioning. Robust tail prediction therefore motivates combining quantile-based models with extreme value theory (EVT).
{Previous works have extended modeling approaches to capture both typical and extreme events by incorporating extreme conditional quantiles through EVT-based methodologies in energy-related forecasting tasks.} {However, these efforts have primarily focused on continuous response variables} \citep{gonccalves2021forecasting,browell2021probabilistic}. For instance, 
\cite{browell2021probabilistic} model electricity load using a mixture of additive QR for non-extreme quantiles, while extreme quantiles are derived from EVT distributions, both conditioned on numerical weather predictions.

Motivated by the need to accurately forecast both extreme and non-extreme fault counts, we propose \textit{X-flexForecast}, a probabilistic framework for modeling fault occurrences across their full range, with particular emphasis on extremes. Non-extreme faults are modeled by linearly interpolating multiple additive quantile regressions, while extremes are captured using a discrete generalized Pareto distribution. Weather-related uncertainty is incorporated via \gls{nwp}, improving reliability in weather-driven forecasts. {The framework targets lead times of up to four days, balancing forecast uncertainty with operational relevance for control-room decision-making.} 

{The remainder of the article is organised as follows. Section~\ref{sec:background} reviews the methodological components underlying \textit{X-flexForecast}, while Section~\ref{sec:methods} details the proposed framework, combining ensemble numerical weather predictions, additive quantile models, and extreme value theory for probabilistic forecasting of discrete outcomes. }Section~\ref{sec:simulation} presents simulation studies based on real data, Section~\ref{sec::application} applies the methodology to the \gls{spen} fault data, and Section~\ref{sec:conclusion} concludes with a discussion of results, limitations, and future directions.

\section{Background}\label{sec:background}

In this section, we provide the {background} for the modeling and forecasting framework. The notation is consistent with the real-data application in Section~\ref{sec::application}. {The data consist of district-level fault counts $y_{d,t}$ at time $t$, where $d$ indexes the districts}. The covariate vector $\mathbf{x}_{d,t}$ contains summary statistics of weather-related variables, and its components can differ across districts as faults occurrence may be driven by different environmental factors. Our aim is to treat $Y_{d,t}$ as a random variable and select suitable models to estimate the conditional distribution $F(Y_{d,t}\given \mathbf{x}_{d,t})$.

\subsection{Quantile Regression} \label{sec:quantilereg}
To avoid assuming a parametric form for $F(Y_{d,t}\mid\mathbf{x}_{d,t})$ and allow smooth dependence on $\mathbf{x}_{d,t}$, we consider multiple quantile regression using generalized additive models following \cite{fasiolo2021fast}. The conditional distribution $F(Y_{d,t}\mid\mathbf{x}_{d,t})$ is constructed from multiple conditional quantiles
\begin{equation}
    q_\alpha(Y_{d,t}\mid\mathbf{x}_{d,t}) = F^{-1}_{Y_{d,t}}(\alpha\mid\mathbf{x}_{d,t}) = g^{-1}\left(
        \beta_{\alpha,0} + \bm{\beta}_{\alpha,A}^\top \mathbf{A}_{d,t} + \sum_{k} f_{\alpha,k}({\bm x}^{S_k}_{d,t})
    \right),
    \label{eq:qgam}
\end{equation}
at probability levels $0<\alpha_1 < \alpha_2 < \dots <1$, where $g(\cdot)$ is a monotone function. The vector $\mathbf{A}_{d,t}$ is a column vector containing the subset of covariates from the full covariate vector $\mathbf{x}_{d,t}$ that enter the model linearly, that is, without being expanded through basis functions. These covariates are directly multiplied by the corresponding regression coefficient vector $\bm{\beta}_{\alpha,A}$, while the remaining covariates enter the model through smooth functions $f_{\alpha,k}(\cdot)$.

Besides the linear terms, we include smooth components indexed by $k$, each defined on a subvector of the covariate vector. Let $S_k \subseteq \{1,\dots,p\}$ denote an index set specifying which components of $\mathbf{x}_{d,t}$ enter the $k$th smooth term. We write $\mathbf{x}_{d,t}^{S_k} = (x_{d,t}^{\{i\}})_{i \in S_k}$. For example, if $S_k = \{2,4\}$, then $\mathbf{x}_{d,t}^{S_k} = (x_{d,t}^{\{2\}}, x_{d,t}^{\{4\}})$. The additive predictor in Equation~\eqref{eq:qgam} includes the $\sum_k f_{\alpha,k}(\mathbf{x}_{d,t}^{S_k})$. Each smooth function $f_{\alpha,k}$ is represented as a linear combination of pre-specified spline basis functions,$
f_{\alpha,k}(\mathbf{x}_{d,t}^{S_k}) = \sum_{j=1}^{K_k} b_{k,j}(\mathbf{x}_{d,t}^{S_k})\, \gamma_{\alpha,k,j},$ where $b_{k,j}(\cdot)$ are basis functions calculated using $\mathbf{x}_{d,t}^{S_k}$, $\gamma_{\alpha,k,j}$ are the associated coefficients, and $K_k$ denotes the number of basis functions.

Optimal smoothing of the quantile (or Pinball) loss is described fully in \cite{fasiolo2021fast} and briefly summarised here. For probability level $\alpha$ the smoothed loss is given by $
\tilde{\rho}_\alpha (y - \mu) = (\alpha - 1) \frac{y - \mu}{\sigma} + \lambda \log \left( 1 + e^{\frac{y-\mu}{\lambda\sigma}}\right)$
where $\lambda>0$ and the usual Pinball loss is recovered when $\lambda \to 0$. Furthermore, the optimal bandwidth $h=\lambda \sigma(\mathbf{x})$ depends on $\mathbf{x}$ for heteroscedastic $y\given\mathbf{x}$, i.e. when the variance of $y$ depends on $\mathbf{x}$. Prior to estimating the regression coefficients, $\mu(\mathbf{x})$ and $\sigma(\mathbf{x})$ are estimated, the latter via a generalized additive model approach (GAM; \citealp{hastie1990gam}). 

Accounting for the discrete nature of $Y_{d,t}$, although QR is traditionally formulated for continuous responses, it can be extended to handle discrete outcomes. \cite{machado2005quantiles} proposed a \emph{jittering}-based approach, which adds uniform random noise to the count variable, effectively transforming it into a continuous variable. This allows standard asymptotic results—such as consistency and asymptotic normality—to hold, making the method particularly useful when inference on model parameters is the primary objective.



\subsection{Discrete generalized Pareto distribution}\label{sec:degpd}

QR has limitations in estimating extreme conditional quantiles due to data sparsity in the extremes, a challenge further exacerbated when conditioning on exogenous variables \citep{koenker2005quantile}, thus, alternative models are required to better characterize {the tail behavior}. 
Studying the right tail of a distribution essentially involves examining exceedances above a sufficiently high threshold $u$. Exceedances are defined as the nonnegative differences between the observed value and the threshold $u$. When the response under study $Y$ is continuous and under max-stability assumptions (see, e.g., \citealp[Ch.~4]{coles_introduction_2001}), these exceedances asymptotically follow a Generalized Pareto (GP) distribution, described by
$ F_{\text{GP}}(y \mid Y > u) = 1 - \left(1 + \frac{\xi(y - u)}{\sigma}\right)^{-1/\xi} $ if $ \xi \neq 0 $, and $ F_{\text{GP}}(y \mid Y > u) = 1 - \exp\left(-\frac{y - u}{\sigma}\right) $ if $ \xi = 0 $. Here $\sigma>0$ and $\xi\in\mathbb{R}$ denote the scale and shape parameters, respectively.
Clearly, the above distribution is defined for $ y > u $, and note that if $ \xi < 0 $, the support of $Y\given Y>u$ is $\{z\in\mathbb{R}:u < z \leq u - \sigma / \xi \}$. 

{
Early regression formulations for threshold exceedances brought the GP into generalized linear and additive modeling settings \citep[e.g.,][]{davison_models_1990, chavez2005generalized}, allowing the scale and shape parameters to vary smoothly with covariates, including those representing complex non-stationary effects such as time or spatial location, provided that sufficient exceedances exist. However, the classical GP assumes a continuous response. To accommodate discreteness, \citet{shimura2012discretization} and \citet{hitz2024extremes} introduced the DGP distribution, which extends GP tail modeling to count-valued outcomes. The DGP has since been applied to diverse discrete-tail problems, including road accidents, disease transmission, and hospital congestion \citep{prieto2014modelling, daouia2023extreme, ranjbar2022modelling}.} Let $Y_{d,t}$ denote a discrete random variable representing the number of faults observed in district $d$ at time $t$, and let $u_{d,t}$ be a high integer threshold. We focus on threshold exceedances, defined as observations satisfying $Y_{d,t} \ge u_{d,t}$. To model tail behavior for discrete data, we assume that the conditional distribution of the threshold exceedances can be represented through a discrete generalized Pareto (DGP) distribution applied to the shifted variable $Y_{d,t} - u_{d,t}$.

Specifically, the conditional cumulative distribution function is defined as
\begin{equation}
F_{\text{DGP}}(y \mid Y_{d,t} \ge u_{d,t})
=
\Pr(Y_{d,t} - u_{d,t} \le y \mid Y_{d,t} \ge u_{d,t})
=
F_{\text{GP}}(y+1) - F_{\text{GP}}(y),
\label{eq:dgp_cdf}
\end{equation}
for integer $y \ge 0$, where $F_{\text{GP}}(\cdot)$ denotes the generalized Pareto distribution with scale parameter $\sigma>0$ and shape parameter $\xi$.

Under this formulation, the support of $Y_{d,t} - u_{d,t} \mid Y_{d,t} \ge u_{d,t}$ is $\{0,1,2,\dots\}$ if $\xi \ge 0$ and $\{0,1,2,\dots,\lfloor -\sigma/\xi \rfloor\}$ if $\xi < 0$, where $\lfloor \cdot \rfloor$ denotes the floor operator. A closed-form expression for the associated probability mass function follows directly from Equation~\eqref{eq:dgp_cdf}, enabling exact likelihood-based inference for the tail model via maximum likelihood estimation. This formulation naturally fits within a generalized additive model for location, scale, and shape (GAMLSS) framework, in which the scale parameter of the tail distribution is modeled as a smooth function of weather predictors $\mathbf{x}_{d,t}$. Estimation is carried out using the GAM-based methods described in \cite{wood2016smoothing}.

\subsection{Bayesian Model Averaging for Weather and Fault Forecasting} \label{sec:bma}

BMA has been extended to dynamical forecast models in \cite{raftery2005using}, where each member of an ensemble forecasting system is represented by a probabilistic distribution. These distributions are combined in a weighted mixture to produce a single density forecast with weights based on past performance of individual members.

{Let $Y$ denote the future fault count to be predicted, and $\mathcal{D}=\{y_1,\dots,y_n\}$ denote the observed fault counts in the training data of length $n$. Consider a collection of $J+1$ forecasting models $\{M_0,\dots,M_J\}$ (with $J=50$ in our application), each yielding an estimated predictive distribution for $Y$ given $\mathcal{D}$. For model $M_j$, let $\hat f_j$ denote a point forecast derived from its estimated predictive distribution (e.g.,  a posterior mean or specific/multiple quantiles). BMA combines the model-specific predictive distributions according to $p(Y \mid \mathcal{D}) = \sum_{j=0}^J p(Y \mid M_j,\mathcal{D})\, p(M_j \mid \mathcal{D}),$
where $p(M_j \mid \mathcal{D})$ is the posterior probability of model $M_j$ given the observed data. The BMA point forecast is then obtained as a functional of the averaged predictive distribution $p(Y \mid \mathcal{D})$, rather than as a weighted average of the individual point forecasts $\hat{f}_j$.}

Here, the fault forecast $\hat{f}_0$ is derived from the single high-resolution deterministic forecast (HRES) member, while $\hat{f}_1,\dots,\hat{f}_{50}$ correspond to the 50-member \gls{ecmwf} ensemble forecast. 
When the $\hat{f}$ comprise multiple predictive quantiles, quantile averaging can be employed for simplicity, although alternative combination strategies, such as the Beta-transformed linear pooling proposed by \cite{ranjan2010combining}
may provide modest improvements in probabilistic calibration. 
{
At very short lead times (approximately 1--2~days), the ensemble is often slightly over-dispersive: the initial perturbations designed to represent atmospheric uncertainty can produce a spread larger than the actual forecast error \citep{leutbecher2009diagnosis}. 
This artificial early spread is corrected within the model to prevent inflated predictive uncertainty. As the forecast horizon increases, the ensemble spread generally becomes more consistent with the observed variability.
 }

{
\section{Methodology} \label{sec:methods}
}
We develop \textit{X-flexForecast}, a probabilistic framework that combines QR for the bulk of the distribution with a DGP model for the tail. This construction links nonparametric conditional quantiles with a parametric tail representation, providing a coherent predictive distribution over the full (discrete) support of $Y_{d,t}$. 

\subsection{Model Specification} \label{sec:model_spec}

The predictive estimator $\widehat{F}_{d,t}(\mathbf{x}_{d,t})$ is defined using the QR formulation in Section~\ref{sec:quantilereg} and the DGP tail component in Section~\ref{sec:degpd}. The QR-based component is used to characterize the conditional distribution up to probability level $\alpha^{(T)}_{d,t}$, while the DGP component models threshold exceedances $ \geq u_{d,t}$. 
For the tail component, we do not adopt a fully non-stationary DGP \citep{chavez2005generalized}. Allowing both scale and shape parameters to vary smoothly over time or space would require substantially more exceedances per series to ensure stable estimation and would markedly increase computational complexity. Instead, we follow a parsimonious specification in which non-stationarity is introduced only through the scale parameter, modeled as $\sigma(\mathbf{x}^{T}_{d,t})$ using covariate-dependent GAMs. Here, $\mathbf{x}^{T}_{d,t}$ denotes the subset of weather predictors included exclusively in the tail model, while the shape parameter $\xi$ is assumed constant.

For the bulk, we estimate conditional quantiles $q_\alpha(y_{d,t}\mid \mathbf{x}_{d,t})$ on a coarse grid of $m$ probability levels $0<\alpha_1<\cdots<\alpha_m<1$, including severe-weather periods where data are sparse. Coefficients are obtained using an optimally smoothed quantile loss, which is robust to sparsity (see Section~\ref{sec:quantilereg}). The predictive distribution $\widehat{F}_{QR}$ is then constructed on the discrete support of $Y_{d,t}$ by linear interpolation between the estimated quantiles. We fit QR directly to the integer responses and avoid jittering or ad hoc rounding, since adding random perturbation can increase bias, reduce interpretability, and incur additional computational cost \citep{machado2005quantiles}.

\subsection{Estimation and Forecast} \label{sec:flexForecast}

Because the response $Y_{d,t}$ is discrete, the additive QR–based quantile $q_{\alpha^{(T)}_{d,t}}(\mathbf{x}_{d,t})$ is generally not an element of the integer-valued support of $Y_{d,t}$. To respect discreteness, we estimate multiple quantiles over discrete probability levels and linearly interpolate up to
$\tilde{\alpha}^{(T)}_{d,t}(\mathbf{x}_{d,t})
= \widehat{F}_{QR}\!\left(\big\lfloor q_{\alpha^{(T)}_{d,t}}(\mathbf{x}_{d,t}) \big\rfloor\right)$,
which defines the predictive distribution for $y_{d,t}\in\big[0,\lfloor q_{\alpha^{(T)}_{d,t}}(\mathbf{x}_{d,t}) \rfloor\big]$. The remaining support, $y_{d,t}\in\big[\lceil q_{\alpha^{(T)}_{d,t}}(\mathbf{x}_{d,t}) \rceil,+\infty\big)$, is modeled by the DGP.

The resulting predictive distribution $\widehat{F}_{d,t}(\mathbf{x}_{d,t})$ is given by
\begin{equation}\begin{cases} 
    \widehat{F}_{{QR}}(y_{d,t} \given \mathbf{x}_{d,t}), & \text{if } y \leq \lfloor q_{\alpha^{(T)}_{d,t}}(\mathbf{x}_{t}) \rfloor, \\
    \tilde{\alpha}^{(T)}_{d,t}(\mathbf{x}_{d,t}) + \left( 1-\tilde{\alpha}^{(T)}_{d,t}(\mathbf{x}_{d,t}) \right)F_{DGP}\left(y- \lceil q_{\alpha^{(T)}_{d,t}}(\mathbf{x}_{d,t}) \rceil \given {{\xi},\sigma(\mathbf{x}_{d,t})} \right), & \text{if } y \geq \lceil q_{\alpha^{(T)}_{d,t}}(\mathbf{x}_{d,t}) \rceil , 
\end{cases}
\label{eq:mixture}
\end{equation}
where $F_{DGP}$ is the CDF of the DGP described in Section~\ref{sec:degpd}, and $\widehat{F}_{QR}$ denotes the values obtained through the linear interpolation described in Section~\ref{sec:model_spec}.

Because the DGP is fitted to exceedances above $\lceil q_{\alpha^{(T)}_{d,t}}(\mathbf{x}_{d,t}) \rceil$, the choice of $\alpha^{(T)}_{d,t}$ is critical. While threshold stability diagnostics, such as assessing the stability of $\hat{\xi}$ across thresholds, are commonly used, we select $\alpha^{(T)}_{d,t}$ empirically by choosing the one with best out-of-sample predictive performance. The selected level must also be sufficiently high for the DGP approximation to remain valid (see Appendix~\ref{app:sensibility_threshold} for sensitivity analyses and Appendix~\ref{app:flex_model_specifications} for the choice details).

{To produce $k$-time-horizon ahead forecasts, ensemble \gls{nwp} outputs provide $j$ forecasts of $\hat{\mathbf{x}}^{(j)}_{d,t\mid t+k}$ available at time $t$ . Passing each member through the fitted model yields member-specific predictive distributions $\widehat{F}(y_{d,t+k}\mid\hat{\mathbf{x}}^{(j)}_{d,t\mid t+k})$ which we combine via BMA (Section~\ref{sec:bma}). To mitigate early over-dispersion and improve short-horizon reliability, we apply a weighted quantile-averaging scheme that emphasizes the HRES/control member at short lead-times. At lead-time 0~h, the HRES/control member is assigned weight 50, while each of the remaining fifty members receives weight 1. The weight assigned to the HRES/control member decreases linearly with lead-time until 72~h, at which point it equals the weights of the other members. Beyond 72~h, all members are equally weighted.}

\section{Implementation \& Benchmarking}\label{sec:simulation}

In this section, we conduct simulation experiments to emulate the data-generating process observed in the occurrence of faults within Scottish districts. The goal is to evaluate whether the proposed model can provide accurate forecasts for both typical and extreme events. The model is compared against alternative approaches suitable for this task, including an additive quantile regression (hereafter referred to as \textit{qgam}) model \citep{fasiolo2021fast} and the extended discrete generalized Pareto model (EDGP); \citealt{ahmad2024extended}). The \textit{qgam} model is treated as a baseline in our simulations, as one of the objectives is to demonstrate that the proposed model provides improved estimation for the tail while maintaining a distribution-free approach for the bulk of the data. In contrast, the EDGP model, an extension of the DGP framework, is included as a competing alternative. This model accommodates discrete responses and extremes without a predefined threshold by modeling the full support of $Y$. However, it relies on a global parametric distribution, which may lack flexibility to capture variability in real data, if compared with \textit{X-flexForecast}.

All results were generated using the implementations provided in the packages referenced in \cite{fasiolo2021package} and \cite{ahmad2024extended}, with default parameter settings. Our proposed model is implemented in the \textsf{R} package \texttt{xflex4cast}
\footnote{Available at: \url{https://github.com/ddoubleblindpeerreview/xflex4cast}.}
which relies on the \texttt{qgam} \citep{fasiolo2021package} and \texttt{GJRM} \citep{marra2024gjrm} packages to fit a \textit{qgam} and a DGP model using a GAMLSS framework, respectively.

\subsection{Simulation Experiments}
\label{sec:simulations_experiments}

Our simulation studies aim to create a setup where the data below and above a known threshold are generated from two distinct distributions, referred to as the "bulk" and the "tail," respectively. 
This approach accounts for two different underlying processes in each case. 
A comprehensive review of this general methodology in the continuous domain is provided by \cite{dey2016extreme}. 
Extending it to a discrete response, as outlined in \cite{ahmad2024extended}, the distribution of the random variable $Y$ is given by: 
\begin{equation}
P_{\text{dg-dgp}}(Y = k) =
\begin{cases} 
\displaystyle  {P_{\text{dg}}(Y = k \mid\boldsymbol{\theta}_B)} & \text{for } k \leq u - 1 \\ 
{\left(1-P_\text{dg}(Y\geq u \mid \boldsymbol{\theta}_{B})\right)}  \times  P_{\text{dgp}}(Y = k \mid Y \geq u; \boldsymbol{\theta}_T) & \text{for } k \geq u
\end{cases}
\label{eq:sim_dgp}
\end{equation}

The exceedance threshold is defined as $u = q_{Y_{\mathrm{dg}}}(1-\phi)
= \inf\{k \in \mathbb{N}_0 : F_{\mathrm{dg}}(k \mid \boldsymbol{\theta}_B) \ge 1-\phi\},$ where $F_{\mathrm{dg}}(\cdot \mid \boldsymbol{\theta}_B)$ denotes the CDF of the discrete gamma bulk distribution and $q_{Y_{\mathrm{dg}}}(\zeta)$ its $\zeta$-quantile. Because the bulk distribution is discrete, its CDF increases only at integer values, so the threshold cannot be placed between integers. Consequently, $\phi$ specifies the target exceedance probability, and $u$ is the smallest integer at which the bulk CDF reaches or exceeds $1-\phi$. Under this setting, the bulk component $P_{\mathrm{dg}}(Y=k \mid \boldsymbol{\theta}_B)$ is defined for $k \le u-1$, while the tail component $P_{\mathrm{dgp}}(Y=k \mid Y \ge u;\boldsymbol{\theta}_T)$ characterizes exceedances for $k \ge u$. Together, these components define the DG–DGP mixture distribution $P_{\mathrm{dg\text{-}dgp}}$. The parameter sets $\boldsymbol{\theta}_B$ and $\boldsymbol{\theta}_T$ correspond to the distributional parameters governing the bulk and tail components, respectively. For the discrete-gamma bulk, $\boldsymbol{\theta}_B~=~(\kappa, \boldsymbol{\beta}_\lambda)$ comprises the shape parameter $\kappa$ and the regression coefficients vector $\boldsymbol{\beta}_\lambda$ associated with the scale parameter $\lambda(\mathbf{z})$, where $\mathbf{z}$ denotes the vector of covariates used in the simulation. For the DGP tail, $\boldsymbol{\theta}_T = (\xi, \boldsymbol{\beta}_\sigma)$ includes the shape parameter $\xi$ and the regression coefficients vector $\boldsymbol{\beta}_\sigma$ defining the scale parameter $\sigma(\mathbf{z})$.

To evaluate the performance of each model, the root mean squared error (RMSE) was calculated for the estimated quantile $\hat{q}_{\alpha}(\mathbf{x}_{d,t})$ against its true value. This was done within a range of probability levels $\alpha = \{0.05, 0.25, 0.5, 0.75, 0.9, 0.95, 0.99, 0.999, 0.9999\}$, with the higher values specifically chosen to emulate the occurrence of extreme events. The experiment was replicated 1000 times for each simulation setting described below, where each setting reflects potential behaviors observed in the occurrence of faults across Scottish districts. To accommodate cases where heavy-tailed behavior may not be consistently observed, the shape parameter $\xi \in \boldsymbol{\theta}_{T}$ is varied over the set $\xi \in \{0.0, 0.03\}$, allowing for flexibility in characterizing the tail behavior. The scenario $\xi < 0$ was omitted from the simulation experiments, as it did not occur in the empirical data from the studied districts, and thus was not considered within the scope of the simulation study, for brevity sake.

The sampling procedure first draws $\mathbf{z}$ from the empirical distribution of observed weather predictors $\mathbf{x}^{T_{k}}_{d,t}$ in {a subset $T_{k}$ of predictors}. {Given $\mathbf{z}$, a draw $Y$ is then generated from the distribution defined in Equation~\eqref{eq:sim_dgp}}. The parameters $\boldsymbol{\theta}_B$ and $\boldsymbol{\theta}_T$ vary with $\mathbf{z}$ according to each simulation scenario. Two simulation exercises are described below to show the model flexibility to capture the complex underlying process under correct specification. Additional experiments which include a misspecified model are provided in Appendix \ref{app:sim_three_scenario}. {For the \textit{X-flexForecast} approach the choice of the quantile-level $\alpha^{(T)}$ is set to $1-\phi$, which corresponds to the true data-generating process defined in Equation~\eqref{eq:sim_dgp}. For further discussion on the sensitivity analysis of this parameter, as well as suitable choices under practical settings where the true $\phi$ is unknown, see Appendix~\ref{app:sensibility_threshold}.}

\subsubsection{First scenario: scale-invariance in the tail distribution}\label{sec:sim_one}

For the first simulation scenario, we define the scale parameter of the discrete gamma distribution, which characterizes the bulk of the distribution conditional on a predictor $\mathbf{z}^{\{1\}}$. The parameters are chosen as $\boldsymbol{\theta}_{B} = \left\{\alpha = 1.5, \gamma(\mathbf{z}^{\{1\}}) = \exp(\beta_{0} + \beta_{1}\mathbf{z}^{\{1\}})\right\}$. The predictor $\mathbf{z}^{\{1\}}$ is sampled from an empirical distribution derived from the \gls{ws10max} values recorded at 24-hour intervals in one of the $d$ Scottish districts. This ensures that the simulated data reflects the distribution of real-world predictors. The true values for the coefficients are set to $\beta_{0} = 1$ and $\beta_{1} = 2$. For the tail of the model, we assume $P_{\text{dgp}} \sim \text{DGP}\left(\xi \in \{0.0, 0.3\}, \sigma = 2.5\right)$, representing light-tailed and heavy-tailed behaviors, respectively. Lastly, the proportion of observations above the threshold is set as $\phi=0.1$. 

For the model specification of the \emph{qgam} approach and the bulk of the \textit{X-flexForecast}, conditional quantiles are modeled using penalized spline smooth functions with a basis dimension of $k=10$. This approach is adopted because the quantile function of the discrete gamma distribution does not have a closed-form expression, and the flexibility of penalized splines accommodates this limitation. The tail model is correctly specified with only constant terms, which are still estimated. For the EDGP model, parameterized by $\kappa$, $\sigma$, and $\xi$, we follow \cite{ahmad2024extended}'s recommendations, employing constant models for $\kappa$ and $\xi$, while $\sigma(\mathbf{z}^{\{1\}})$ is estimated via a penalized spline with $k=10$, ensuring consistency with the QR approach.

The results are summarized in Table \ref{tab:sim_scenario_one}, which presents the average RMSE calculated over 1000 replications of the simulated experiment. For readability, quantiles below the probability level of $\alpha = 0.95$ are omitted in the second and fifth columns, as the \textit{X-flexForecast} and \textit{qgam} models yield identical results in this range. The $\alpha = 0.9$ quantile is also excluded for brevity. The results indicate that for most bulk quantiles, the EDGP model achieves the smallest RMSE values, although these are nearly on the same order of magnitude as those of the alternative methods. {For the bulk quantiles, this likely occurs because EDGP benefits from the joint estimation of quantiles implied by its parametric distributional assumption, whereas the alternative methods estimate these quantiles independently. As the probability level associated with the quantiles increases, the \textit{X-flexForecast} model consistently achieves notably lower RMSE values relative to competing approaches, reflecting the advantage in tail predictive performance afforded by its greater flexibility. This demonstrates its capacity to provide accurate estimation across both the bulk and tail regions of the distribution, emphasizing its adaptability in capturing the entire predictive range more effectively than alternative methods.}


\begin{table}[ht]
    \vspace{-0.75cm}
    \centering
    \small
    \resizebox{\textwidth}{!}{
    \begin{tabular}{c|cccc|cccc}
        \multicolumn{9}{c}{} \\
        \multicolumn{1}{c}{} & \multicolumn{4}{c}{$\xi = 0.0$} & \multicolumn{4}{c}{$\xi = 0.3$} \\
        \hline
        $\alpha$ & $\bar{q}(\mathbf{X})$ &\textit{ X-flexForecast} & \textit{qgam} & EDGP & $\bar{q}(\mathbf{X})$ & \textit{X-flexForecast} & \textit{qgam} & EDGP \\
        \hline
        &  & \multicolumn{3}{c|}{RMSE} &  & \multicolumn{3}{c}{RMSE} \\
        \hline
        0.05   & 0.09 & 0.60 $\pm$ 0.32 & --- & \textbf{0.52 $\pm$ 0.06} & 0.09 & 0.60 $\pm$ 0.34 & --- & \textbf{0.55 $\pm$ 0.12} \\
        0.25   & 1.41 & 1.23 $\pm$ 0.40 & --- & \textbf{0.96 $\pm$ 0.05} & 1.42 & 1.22 $\pm$ 0.33 & --- & \textbf{0.98 $\pm$ 0.42} \\
        0.50   & 3.98 & 1.53 $\pm$ 0.50 & --- & \textbf{1.01 $\pm$ 0.11} & 3.99 & 1.53 $\pm$ 0.54 & --- & \textbf{1.04 $\pm$ 0.91} \\
        0.75   & 8.03 & 1.94 $\pm$ 0.66 & --- & \textbf{1.17 $\pm$ 0.30} & 8.04 & 1.93 $\pm$ 0.68 & --- & \textbf{1.23 $\pm$ 1.71} \\
        0.95   & 14.52 & \textbf{2.03 $\pm$ 1.13} & 2.20 $\pm$ 0.77 & 3.26 $\pm$ 0.76 & 14.98 & \textbf{2.09 $\pm$ 1.07} & 2.30 $\pm$ 0.67 & 4.07 $\pm$ 3.63 \\
        0.99   & 18.55 & \textbf{2.14 $\pm$ 1.18} & 3.52 $\pm$ 0.63 & 11.17 $\pm$ 2.16 & 21.53 & \textbf{2.50 $\pm$ 1.10} & 3.91 $\pm$ 0.94 & 12.81 $\pm$ 5.93 \\
        0.999  & 24.27 & \textbf{2.69 $\pm$ 1.62} & 14.18 $\pm$ 4.06 & 22.68 $\pm$ 4.26 & 38.84 & \textbf{4.36 $\pm$ 2.65} & 12.34 $\pm$ 5.63 & 25.10 $\pm$ 9.77 \\
        0.9999 & 30.13 & \textbf{3.91 $\pm$ 2.81} & 60.07 $\pm$ 15.27 & 34.13 $\pm$ 6.39 & 73.33 & \textbf{11.99 $\pm$ 9.72} & 40.33 $\pm$ 20.18 & 40.29 $\pm$ 14.03 \\
        \hline
    \end{tabular}}
    \caption{Average RMSE over 1000 replications for the first simulation scenario. Columns \textit{X-flexForecast}, \textit{qgam}, and EDGP show RMSEs per method; columns $\bar{q}(\mathbf{X})$ report average true quantiles at level $\alpha$. Bold indicates lowest RMSE. Bulk quantiles for \textit{qgam} are omitted as they match those of \textit{X-flexForecast}.}
    \label{tab:sim_scenario_one}
\end{table}

\subsubsection{Second scenario: non-stationary in the tail distribution}\label{sec:sim_two}

{In the second scenario, tail non-stationarity is introduced through covariate dependence of the DGP scale parameter, by allowing $\sigma$ to vary as a function of an additional weather covariate $\mathbf{z}^{\{2\}}$. This notion of non-stationarity is purely covariate-driven and does not involve spatial or temporal dependence.} This setup mimics real-world patterns observed in extreme event occurrences associated with fault counts. The predictor set governing the distribution of $Y$ is now defined as $\mathbf{Z} = \{\mathbf{z}^{\{1\}},\mathbf{z}^{\{2\}}\}$, where both covariates are sampled from a joint empirical distribution estimated from observed data. 
While $\mathbf{z}^{\{1\}}$ continues to represent \gls{ws10max}, $\mathbf{z}^{\{2\}}$ denotes the 90th quantile of total precipitation measured over 24-hour intervals in the same Scottish district. Given their correlation, the two variables are sampled jointly using their empirical joint distribution.
The scale parameter $\sigma$ from the tail is modeled as a function of $\mathbf{z}^{\{2\}}$ via the relationship $\sigma(\mathbf{z}^{\{2\}}) = \exp\left(\beta_{0} + \beta_{1}\mathbf{z}^{\{2\}}\right)$, with parameter values $\beta_{0} = -0.05$ and $\beta_{1} = 0.85$. These values were obtained by fitting the same DGP model to threshold exceedances in real 24-hour data from one Scottish district, using maximum likelihood estimation. The aim was to closely replicate the characteristics of the dataset analysed in  Section~\ref{sec::application}.
All remaining parameters governing the discrete gamma distribution remain unchanged.

The model specification for the \textit{qgam} is modified to include a set of basis functions of size $k=10$ for $\mathbf{z}^{\{2\}}$, and the same modification is applied to the scale parameter of the EGPD. The results are summarized in Table \ref{tab:sim_scenario_two}. The RMSE results remain consistent with those reported in Section~\ref{tab:sim_scenario_one}, indicating comparable performance between \textit{X-flexForecast} and alternative methods for estimating bulk quantiles. However, for higher quantiles, \textit{X-flexForecast} continues to demonstrate superior predictive accuracy, as reflected in the lower average RMSE values.

\begin{table}[htb]
    \vspace{-0.75cm}
    \centering
    \small
    \resizebox{\textwidth}{!}{
    \begin{tabular}{c|cccc|cccc}
        \multicolumn{9}{c}{} \\
        \multicolumn{1}{c}{} & \multicolumn{4}{c}{$\xi = 0.0$} & \multicolumn{4}{c}{$\xi = 0.3$} \\
        \hline
        $\alpha$ & $\bar{q}(\mathbf{X})$ & \textit{X-flexForecast} & \textit{qgam} & EDGP & $\bar{q}(\mathbf{X})$ & \textit{X-flexForecast} & \textit{qgam} & EDGP \\
        \hline
        &  & \multicolumn{3}{c|}{RMSE} &  & \multicolumn{3}{c}{RMSE} \\
        \hline
        0.05   & 0.09 & \textbf{0.60 $\pm$ 0.33} & --- & 0.62 $\pm$ 0.08 & 0.09 & 0.61 $\pm$ 0.36 & --- & \textbf{0.58 $\pm$ 0.08} \\
        0.25   & 1.41 & 1.21 $\pm$ 0.29 & --- & \textbf{0.99 $\pm$ 0.07} & 1.41 & 1.22 $\pm$ 0.35 & --- & \textbf{0.98 $\pm$ 0.08} \\
        0.50   & 3.99 & 1.53 $\pm$ 0.52 & --- & \textbf{0.95 $\pm$ 0.10} & 3.98 & 1.53 $\pm$ 0.53 & --- & \textbf{0.98 $\pm$ 0.16} \\
        0.75   & 8.04 & 1.95 $\pm$ 0.64 & --- & \textbf{1.05 $\pm$ 0.30} & 8.03 & 1.91 $\pm$ 0.65 & --- & \textbf{1.11 $\pm$ 0.37} \\
        0.95   & 13.31 & \textbf{1.94 $\pm$ 1.08} & 2.30 $\pm$ 0.62 & 3.55 $\pm$ 0.71 & 13.52 & \textbf{2.00 $\pm$ 1.06} & 2.39 $\pm$ 0.62 & 3.67 $\pm$ 0.79 \\
        0.99   & 15.34 & \textbf{2.10 $\pm$ 1.09} & 4.44 $\pm$ 0.78 & 12.14 $\pm$ 2.01 & 16.63 & \textbf{2.37 $\pm$ 1.13} & 4.50 $\pm$ 0.86 & 11.75 $\pm$ 2.07 \\
        0.999  & 17.90 & \textbf{2.72 $\pm$ 1.40} & 18.11 $\pm$ 4.43 & 24.93 $\pm$ 4.00 & 24.56 & \textbf{3.89 $\pm$ 2.27} & 15.13 $\pm$ 4.60 & 22.23 $\pm$ 4.03 \\
        0.9999 & 20.55 & \textbf{4.01 $\pm$ 2.68} & 64.54 $\pm$ 15.42 & 37.69 $\pm$ 6.02 & 40.41 & \textbf{9.38 $\pm$ 6.06} & 51.87 $\pm$ 16.12 & 31.72 $\pm$ 5.45 \\
        \hline
    \end{tabular}}
    \caption{Average RMSE over 1000 replications for the second simulation scenario. Columns \textit{X-flexForecast}, \textit{qgam}, and EDGP show RMSEs per method; columns $\bar{q}(\mathbf{X})$ report average true quantiles at level $\alpha$. Bold indicates lowest RMSE. Bulk quantiles for \textit{qgam} are omitted as they match those of \textit{X-flexForecast}.}
    \label{tab:sim_scenario_two}
\end{table}

\section{Case Study: short-term fault forecast for Scotland's electrical networks}\label{sec::application}
The objective of our new flexible forecasting approach is to provide short-term fault forecasts tailored to end-users. 
Specifically, it aims to generate probabilistic fault forecasts for individual districts---organizational units of approximately 1000 km$^2$---within electricity distribution networks.
These forecasts extend up to 4 days ahead, with a 24-hour resolution. This approach explicitly accounts for the heavy-tailed behavior that may occur in fault distributions. The specification aligns with the needs of end-users, who allocate resources from within a day to a few days ahead and require robust risk management, needing accurate uncertainty quantification. 

\subsection{Data description}~\label{sec:DataDescription}
Faults have many causes, not just severe weather; however, periods with large numbers of  faults usually correspond to severe weather, including storms, snow and ice, and extreme heat.  {To address this, we combine detailed network data with high-resolution observed and forecast weather variables to build eleven district-level models for fault forecasting across the two licence areas managed by \gls{spen}, which are located in two distinct regions of the UK, using a daily temporal resolution from March 2010 to April 2024.}

\subsubsection{Network and Fault Data}~\label{sec:DataDescription_network}
\begin{figure}
   \centering
   \includegraphics[width=0.5\linewidth]{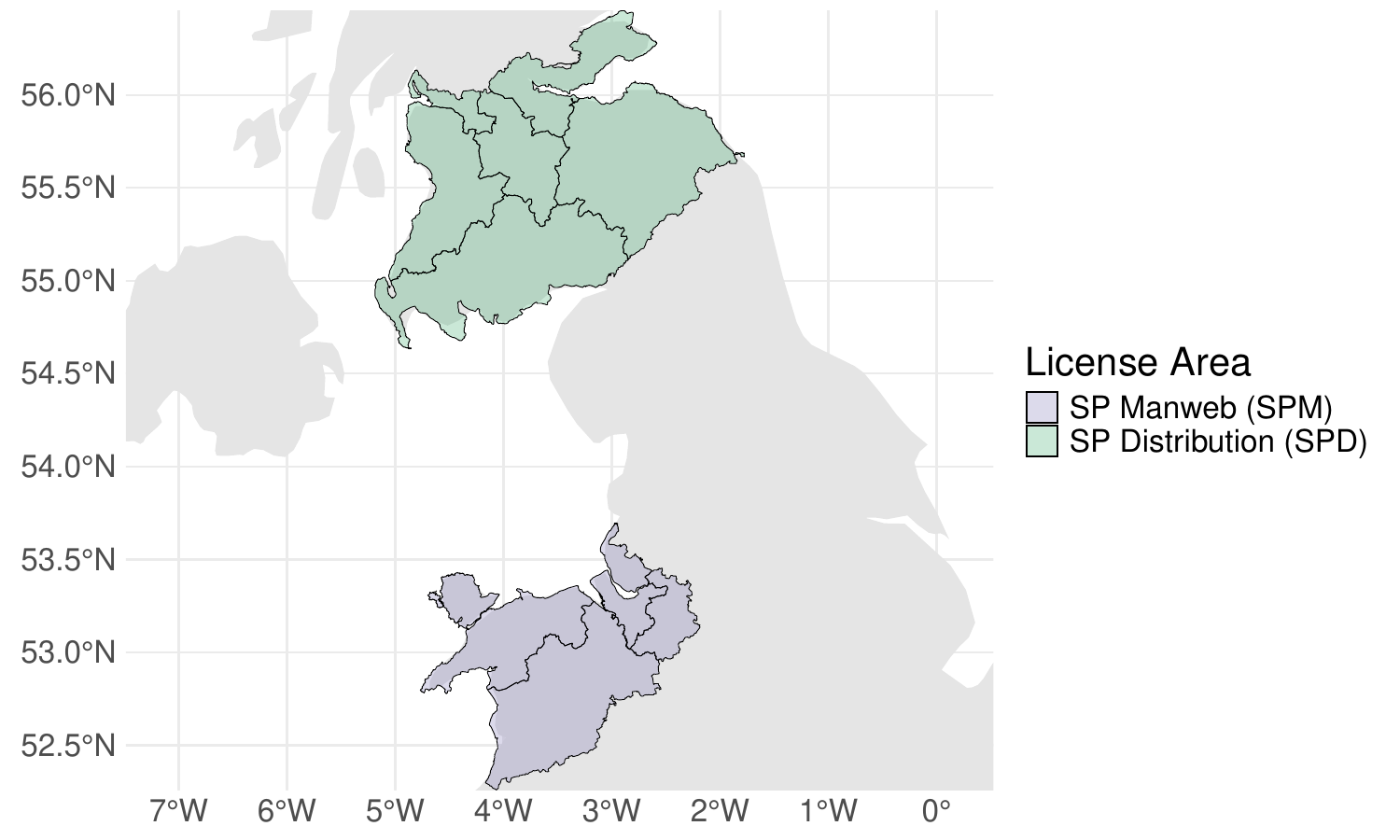}
   \caption{UK map of the two licence areas operated by \gls{spen}: SP Distribution (SPD), covering Central and Southern Scotland (top, in green), and SP Manweb (SPM), covering North Wales, Merseyside, and Cheshire (bottom, in purple). {District boundaries - spatial units over which faults are aggregated for modeling and prediction - are shown within license areas. Each district comprises several hundred circuits.}
   }
   \label{fig:spen_maps}
\end{figure}

Predicting the number of faults in a given time period and {district} is the objective of this work; as such, fault records are required for both modeling and forecast evaluation. Fault recording is a regulatory requirement, so records are well maintained. However, these records are not produced for forecasting purposes and require careful cleaning. A frequent issue is that faults are recorded at the time they are discovered, which is not necessarily the time they occurred. For example, faults caused by a storm may go unnoticed for days, only becoming apparent as repairs are carried out on damaged parts of the network. Time stamps do not accurately record the precise time a fault occurred and must therefore not be over-interpreted. {The most consistent and reliable information is recorded at the \emph{circuit} level (the smallest operational unit), although more precise geo-locations are available only for a minority of events. We aggregate circuit-level fault counts into 24-hour windows and then sum these counts within each district (each district comprises many circuits, typically several hundred). All modeling and prediction are performed at the district level: each observation in our analysis corresponds to a district–day fault count. The 11 districts of \gls{spen}'s distribution network are shown in Figure~\ref{fig:spen_maps}.}

\subsubsection{Weather Data}~\label{sec:DataDescription_weather}
For fault modeling, weather covariates are estimated at an optimized set of locations for each circuit and district using ERA5 reanalysis data \citep{Hersbach2018ERA5Present}, with bilinear interpolation applied to hourly fields. The variables considered include 10-m wind speed and gusts, 2-m temperature and dew point, total precipitation, snow depth, soil moisture, and CAPE. To align with the daily aggregation of fault counts, district-level 24-hour means and maxima are computed. {This temporal and spatial summarization also helps to account for and mitigate potential spatio-temporal dependence at the original hourly and spatial resolution of the weather covariates.} 

Using these aggregated covariates, models are then trained, and fault forecasts are subsequently obtained by applying the fitted models to post-processed numerical weather predictions from the ECMWF high-resolution (HRES) deterministic forecast and the 50-member ensemble forecast. Weather variables are extracted and aggregated consistently across ensemble members and bias-corrected using linear models fitted to historical data, yielding a 51-member ensemble of inputs at daily resolution for lead times up to four days. Final probabilistic fault forecasts are produced via the BMA approach described in Section~\ref{sec:bma}.

\subsubsection{Data Exploration and Combination}~\label{sec:DataDescription_combination}
Network and fault data were explored at the circuit level, but initial modeling experiments revealed that predicting faults at individual circuits was extremely challenging. It was observed that while certain circuits are more vulnerable to faults, due to elevation and vegetation, and experienced more faults overall, during severe weather they experienced similar numbers of faults to other circuits. The high numbers of faults observed at the district level during severe weather are a result of many circuits experiencing small numbers of faults, not vulnerable circuits experiencing many faults. As fault data is particularly sparse for less-vulnerable circuits, modeling the occurrence of these faults was not possible. Therefore, we aim to predict faults at the district level using engineered features (covariates) derived from weather locations corresponding to individual circuits. 

\subsection{Results}\label{sec::results}

The above methodology has been implemented in a case study comprising the 10 of the 11 districts of \gls{spen}'s distribution networks in Great Britain which span Southern Scotland and parts of North-West England and North Wales (see Figure~\ref{fig:spen_maps}). Glasgow \& Clyde North was omitted from this analysis due to low fault counts and therefore uninformative results. Model selection and evaluation are {based on} historical fault data and ERA5 weather reanalysis data from April 2010 to March 2024.
A cross-validation setup with 14 folds is used, where each fold covers a full year from April to March, ensuring that each one includes a single regulatory year, i.e: with one complete winter season, aligning with the operational management of \gls{spen}, and mitigates potential bias from temporal dependence across folds.
Forecasts are evaluated using historical weather forecasts from the ECMWF retrieved from the operational archive in the case of the HRES model, and the TIGGE archive in the case of the \gls{eps}. Historic forecast covers the period April 2018 to March 2024, and include one forecast run (base time) of midnight UTC, and lead-time from 0 to 96h ahead at the maximum temporal resolution available.

For operational use, the forecast number of faults in a given district and time resolution is represented as a categorical variable $c_{d,t} \in \{R, A, G\}$, corresponding to the red, amber, and green (RAG) fault bands, respectively. The thresholds defining the Red-Amber and Amber-Green boundaries vary across districts and time resolution, with $\tau^{(RA)}_{d,t}$ and $\tau^{(AG)}_{d,t}$ representing these transitions for each district and time scale, as risk management strategies differ by location and time interval. The probability of each category for a given observation is determined by calculating $\widehat{F}_d(\tau^{(RA)}_{d,t} \mid \mathbf{x}_{d,t})$ and $\widehat{F}_d(\tau^{(AG)}_{d,t} \mid \mathbf{x}_{d,t})$ using Equation~\eqref{eq:mixture}, which are then used to derive the final probabilities for each fault band.
Additionally, the assignment of a category ${c}_{d,t}$ to an observation $\mathbf{x}_{d,t}$ for communication with end-users is not only based on the most probable category but follows a set of heuristic rules. Specifically, the green category is assigned if its probability exceeds 80\%, the red category is assigned if its probability exceeds 20\%, and the amber category is assigned if neither of the previous conditions holds but its probability is greater than that of red. Fault bands and the logic for assigning categorical predictions were selected though structured conversations and testing with end-users to align with their operational requirements. 

In the \textit{qgam} models, 10-meter wind speed, an indicator variable corresponding to whether trees are in leaf or not, prevailing wind direction, and CAPE were considered candidate weather covariates during model selection. The models incorporated linear terms or cubic spline terms with 10 to 15 basis functions per district to ensure sufficient flexibility in capturing nonlinear relationships. The inclusion of covariates beyond wind speed—namely tree leaves, wind direction, and lightning proxy (CAPE)—for specific districts is consistent with recent studies reporting significant associations between some of these predictors and power outages in the UK \citep{manning2025antecedent}. 

Model selection was guided by out-of-sample (OOS) performance within a $k$-fold cross-validation framework, using pinball loss, sharpness, the Brier score (BS), and the area under the ROC curve (AUC). The BS and AUC computed on the OOS sets were used to select both the covariate set $\mathbf{x}_{d,t}$ entering the DGP tail parameters and the transition probability level $\alpha_{d,t}^{(T)}$, accounting for the role of the categorical variable $\bm{c}_{d,t}$ in forecast usability. Quantiles in the \textit{qgam} component were estimated at probability levels $\boldsymbol{\alpha}=\{0.05,0.25,0.5,\alpha_{d,t}^{(T)}\}$, with $\alpha_{d,t}^{(T)} \in \{0.75,0.8,0.9,0.95\}$. Among competing specifications, preference was given to configurations achieving the largest average improvement in BS and AUC across categories; when the two criteria disagreed, BS was prioritized as the primary selection metric, as it is a strictly proper scoring rule assessing calibration and accuracy, whereas AUC primarily reflects ranking performance \citep{gneiting2007strictly} mostly.
The final variable selections for each tail model, the $\alpha_{d,t}^{(T)}$, and the criteria are summarized in Appendix~\ref{app:flex_model_specifications}.

\subsubsection{Model Evaluation}\label{sec::model_evaluation}

\begin{figure}[ht]
    \centering
    \includegraphics[width=0.6\linewidth]{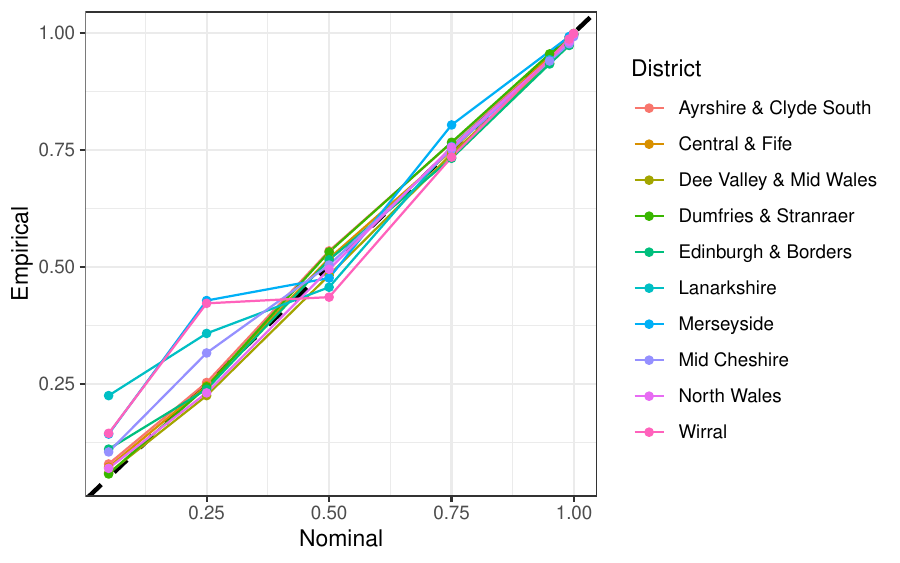}
    \caption{Reliability of \textit{X-flexForecast} models for 24-hour resolution forecast, out-of-sample hindcasts using ERA5 for the windiest 20\% of days.}
    \label{fig:qgam_model_rel}
\end{figure}

The calibration of the fault prediction models is verified by examining Reliability Diagrams for the hindcasts produced in the k-fold cross-validation exercise using on ERA5 reanalysis. Evaluating calibration is challenging due to the integer-valued and right-skewed nature of faults, therefore we consider the conditional reliability for fault predictions when $\mathbf{x}_{d,t}$ includes wind speed in it's highest quintile. These diagrams, one for each district, are shown in Figure \ref{fig:qgam_model_rel} for the \textit{X-flexForecast} models. 
The 24-hour resolution predictions demonstrates good calibration, although some districts with infrequent fault occurrences show poor calibration for the $\alpha_{0.25}$ quantile. 
This is due to the boundary at zero faults, which causes a concentration of probability mass at zero. 

To evaluate the improvement in model performance due to the inclusion of the DGP model within the \textit{X-flexForecast} framework, we computed the ratio of the AUC values between our method and the \textit{qgam} baseline for each district. Additionally, the BS skill score (SS) was calculated using \textit{qgam} as the reference model. 

Figure \ref{fig:24h_tail_evidence} summarize the results for the 24-hour resolution across all districts. In the left panel, AUC ratio values greater than one indicate improved performance, while values less than one indicate worse performance compared to the \textit{qgam}-only approach. In the right panel, values above zero represent an improvement achieved by \textit{X-flexForecast} in predicting the probabilities for each category $C$, while values below zero indicate a decrease in performance. In both panels (left and right), dashed lines represent reference values, indicating no change in performance between the compared approaches. 

{Analysis of the left and right panels reveals that most values lie above the reference line, indicating that incorporating tail models alongside \textit{qgam} improves the estimation of the fault band. For certain districts—such as {Central \& Fife,} Mid Cheshire and Wirral—values below the reference line suggest potential underperformance in the Red category. However, given the low frequency of Red events in these districts, misclassifying some Red events as Amber does not necessarily indicate model failure, as the estimated probability of correct classification may still be informative. {Consequently, relying only on the AUC would be insufficient: the model produces $\widehat{F}_{d,t}(\mathbf{x}_{d,t})$, and their evaluation would require scoring rules also suited to probabilistic forecasting. In the right panel—which reports the SS for the Brier Score (BS)—most districts lie to the right of the reference line, with only some categories in Dee Valley \& Mid Wales, Lanarkshire, and North Wales falling slightly below it. When interpreted jointly with the AUC ratios in the left panel, the overall pattern continues to support the proposed approach. Figure~\ref{fig:24h_tail_evidence} therefore indicates that, on average, even in districts where the fault band is occasionally misclassified, the probabilistic forecasts from \textit{X-flexForecast} are more coherent to the observed categories than those obtained under standard \textit{qgam}. Appendix~\ref{app:flex_model_specifications} provides additional results as threshold-weighted continuous ranked probabilistic score, which also captures the quality of the full probabilistic forecasts, together with the micro and macro AUC values. }

\begin{figure}[!htpb]
    \centering
    \includegraphics[width = \linewidth]{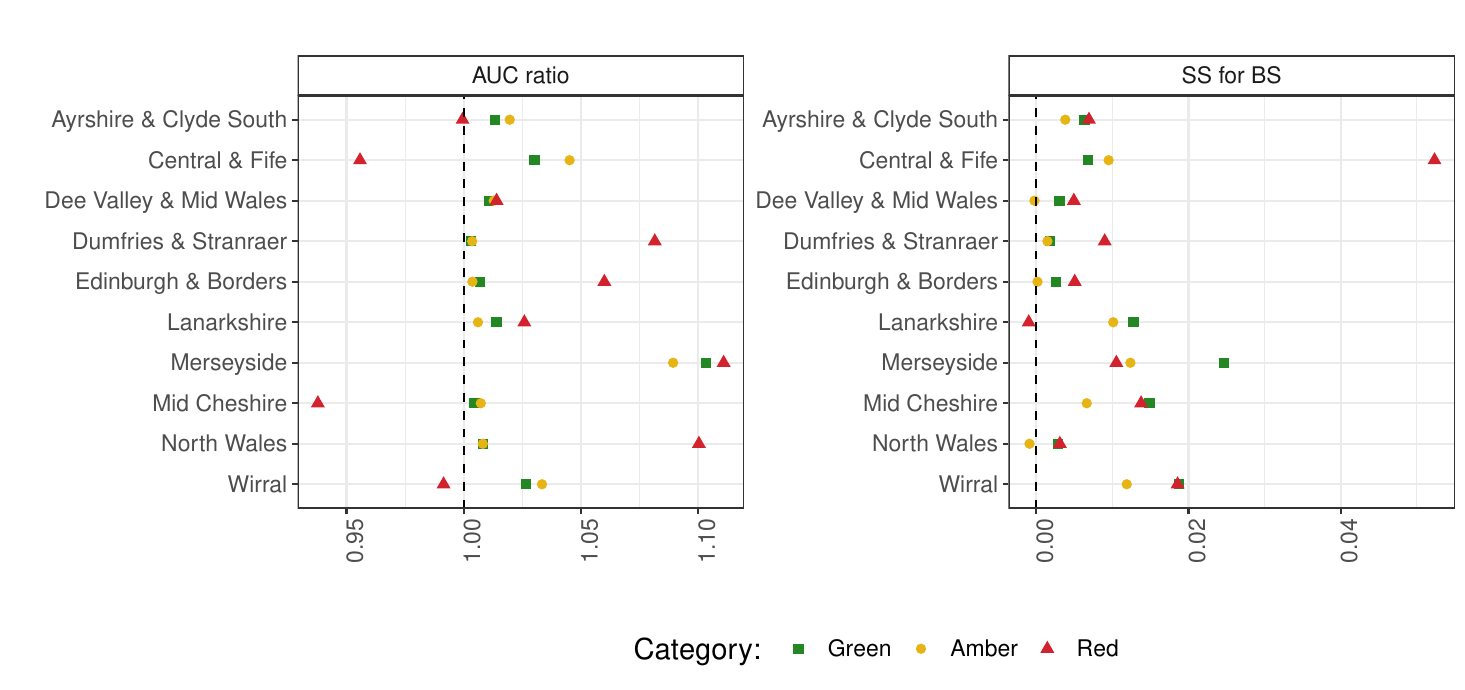}
    \caption{24-hour resolution results across all districts. Left panel: AUC ratio $>1$ indicates improvement over \textit{qgam}-only approach. Right panel: Values $>$0 show \textit{X-flexForecast} improvement in predicting category display probabilities. Dashed lines represent no change in performance.}
    \label{fig:24h_tail_evidence}
\end{figure}

\subsubsection{Forecast evaluation}\label{sec:forecast_evaluation}

Re-forecasts of faults have been produced for the period April 2017 to March 2024 using historic \gls{nwp} and the \textit{X-flexForecast} from the cross-validation exercise trained on data excluding the period being re-forecast. Fault forecasts are produced for each member of the multi-model ensemble, comprising the single HRES member and 50 members of the \gls{eps}, and combined using \gls{bma} and the weighted averaged described in Section~\ref{sec:flexForecast}. For comparison, re-forecasts using the \gls{eps} only and HRES only have also been produced.

The calibration of resulting fault forecasts has been verified, as the models were initially conditional on high wind days. The results for the 24-hour resolution forecasts, shown in Figure \ref{fig:qgam_24h_forecast_rel}, confirm the calibration for most districts and lead times (not shown). The combination of HRES and the \gls{eps} yields improved calibration for most districts.
However, it does not, and was not expected to, address the poor calibration observed in three districts where the fault models had previously shown deficiencies.
\begin{figure}[ht]
    \centering
\includegraphics[width=\textwidth]{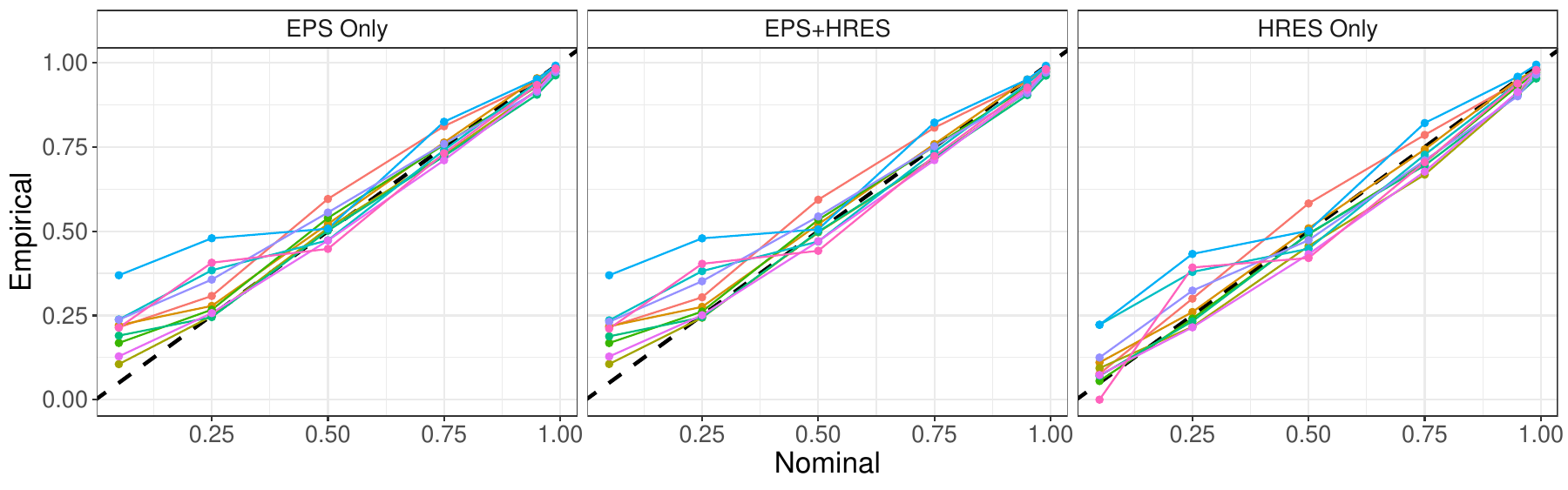}
    \caption{Reliability of \textit{qgam} models, out-of-sample re-forecast with lead-times from zero to four days-ahead using the ECMWF, \gls{eps} and/or HRES deterministic forecast for the windiest 20\% of days. The color for each district is the same as Figure \ref{fig:qgam_model_rel}.}
    \label{fig:qgam_24h_forecast_rel}
\end{figure}

The overall performance of the fault forecast has been evaluated using the Pinball loss for the seven quantiles $\alpha=\{0.05,0.25,0.5,0.75,0.95,0.99,0.999\}$. Figure \ref{fig:qgam_forecast_pinball} shows the scaled Pinball Loss by lead-time and district for 24-hour \textit{X-flexForecast} forecasts. Pinball Scores are scaled per district using the EPS+HRES value at lead-time 0 as baseline. While inter-district comparison is limited due to heterogeneity and low event frequency, the results confirm that forecast quality degrades modestly with lead-time. EPS+HRES consistently outperforms both EPS-only and HRES-only, especially at shorter lead-times (0–48 h), while HRES-only shows the highest losses across most districts, highlighting its reduced reliability when used alone.

\begin{figure}[ht]
    \centering
    \includegraphics[width=0.9\textwidth]{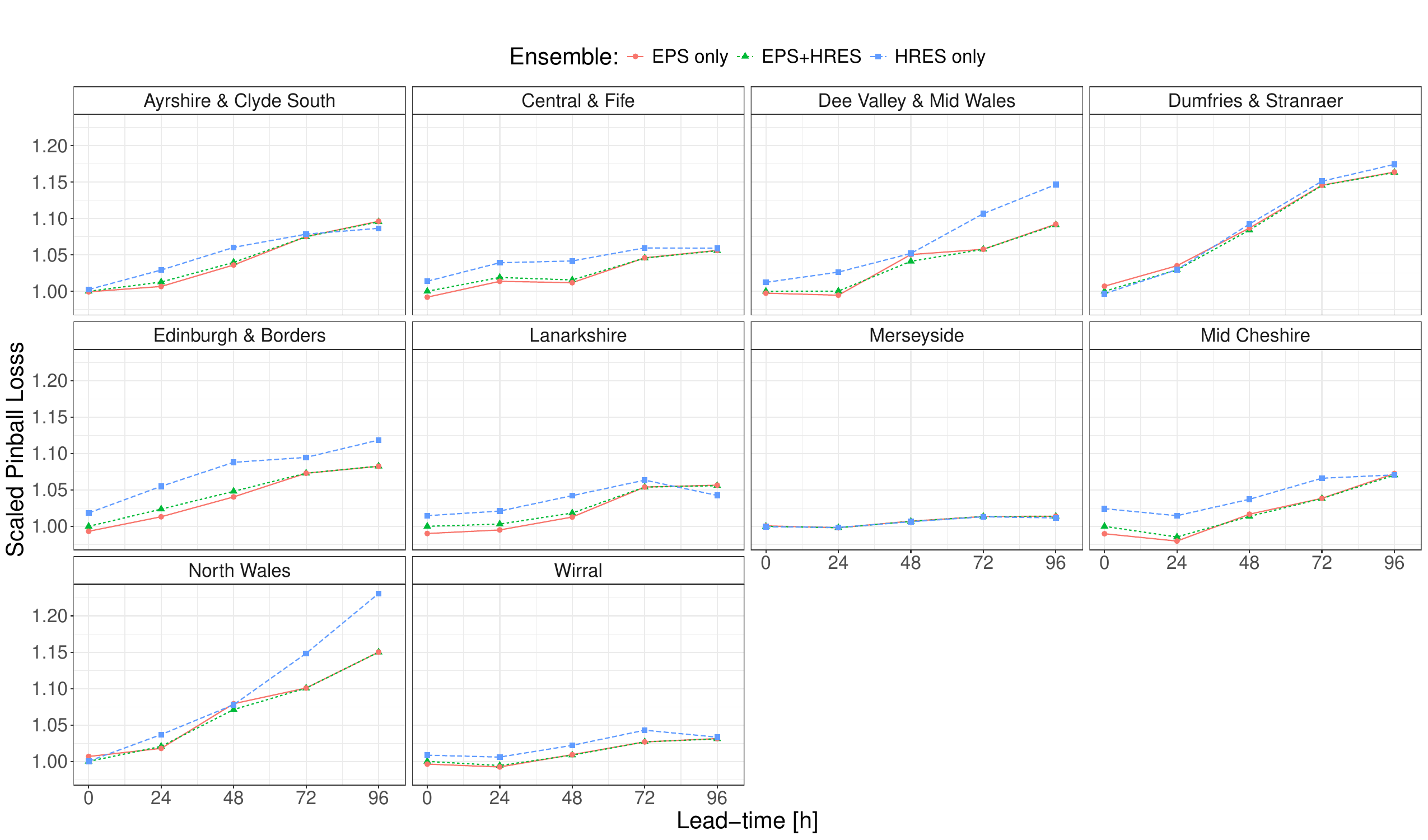}
    \caption{Scaled Pinball Loss for \textit{X-flexForecast} models, based on out-of-sample re-forecasts with lead-times from 0 to 96 hours (24-hour resolution), using EPS and/or HRES forecasts. Scores are scaled per district using the EPS+HRES forecast at lead-time 0 as reference.}
    \label{fig:qgam_forecast_pinball}
\end{figure}

In addition to the above quantitative evaluation, the usability and usefulness of these forecasts have been assessed qualitatively by engineers at \gls{spen}. The forecasting framework has been implemented in a web application developed by Sia and trialed at SPEN from October 2024 to {the time of writing. During this period , engineers reported that tool enhanced situational awareness and operational readiness ahead of severe weather events by providing an additional layer of quantitative evidence to support rapid decision-making under uncertainty. In multiple specific storm-responses, the tool improved coordination and prioritization of network interventions, demonstrating tangible gains in operational efficiency and reducing the duration of power outages.} Overall, the solution was found to be sufficiently accurate to inform operational decision-making and to provide measurable value to both \gls{spen} and electricity consumers relative to business as usual.

\section{Conclusion}\label{sec:conclusion}

Forecasting faults in electricity distribution networks is an essential capability for the resilience of electricity supply. We have developed a fault modeling and forecasting approach that quantifies the two main sources of uncertainty impacting fault prediction: the propensity of severe weather to cause faults, and the future weather conditions. Additionally, the solution provided by \textit{X-flexForecast} is tailored to consider the occurrence of extreme events, and discrete responses. The resulting probabilistic fault forecasts have been verified and evaluated quantitatively and qualitatively. They are found to be of high quality and suitable for operational decision-making.

These results constitute a proof-of-concept forecasting capability, demonstrating the potential for future operational deployment while also indicating avenues for further refinement. One such avenue concerns the selection of the threshold probability level $\alpha^{(T)}_{d,t}(\mathbf{x}_{d,t})$. In this study, the parameter is determined via cross-validation; however, an automated selection strategy, similar in spirit to \cite{murphy2024automated} and adapted to the \textit{X-flexForecast} framework, could reduce the computational burden associated with model estimation and deployment, leading to a more efficient implementation.
Additional extensions are possible but nontrivial and beyond the scope of this work. These include adding covariates not currently used in fault models (e.g., time-varying vegetation), generalising to networks under different climates, expanding the multi-model ensemble, adapting the model to changes in \gls{nwp}, exploring alternative interpolation schemes for constructing $\widehat{F}$ from multiple quantile regressions, and optimizing or replacing BMA-based forecast combination. Future work could also introduce explicit spatio-temporal terms in the additive predictor, including the tail component \citep{chavez2005generalized}, to capture dependence across districts and time, at the cost of increased complexity.

\section{Achnowledgements}

We would like to thank Abdelouahab Mestassi, Jeremy Joslove, Kitti Kovacs, Léopold Duverger-Noinski, Milan Clerc, Miao Zhou, S\'ebastien Gerber, Simon Hessami, Simone de Angelis, Wilbur Zhu from Sia, as well as \gls{spen} and the UK Met Office for their contributions to this work. We also acknowledge the use of ERA5-Land data \citep{Hersbach2018ERA5Present}, ECMWF HRES, and ENS. This work is part of the \textit{Predict4Resilience} project, funded through the Strategic Innovation Fund administered by OFGEM and UKRI. 

The weather data used in study are available from ECMWF: ERA5 is available from the Climate Data Store\footnote{https://cds.climate.copernicus.eu/}, historic ECMWF ensemble forecasts from the TIGGE archive\footnote{https://apps.ecmwf.int/datasets/data/tigge/}, historic ECMWF deterministic/HRES forecasts from the ECMWF operational archive\footnote{https://www.ecmwf.int/en/forecasts/dataset/operational-archive}.  
Restrictions apply to the availability of electricity network fault data, which were used under license for this study. Data are available from the authors with the permission of \gls{spen}. Code to reproduce the simulation experiments presented in Section~\ref{sec:simulations_experiments} is available on GitHub\footnote{Available at \url{https://github.com/ddoubleblindpeerreview/xflex4cast}.} and archived on Zenodo\footnote{DOI: \href{https://doi.org/10.5281/zenodo.18173632}{10.5281/zenodo.18173632}}.

We have no conflicts of interest to declare. For the purpose of open access, the authors have applied a Creative Commons Attribution (CC BY) licence to any Author Accepted Manuscript version arising from this submission.

\bibliographystyle{CUP}
\bibliography{biblio.bib}

@article{cerrai2019storm,
   title={Predicting storm outages through new representations of weather and vegetation},
  author={Cerrai, Diego and Wanik, David W and Bhuiyan, Md Abul Ehsan and Zhang, Xinxuan and Yang, Jaemo and Frediani, Maria EB and Anagnostou, Emmanouil N},
  journal={IEEE Access},
  volume={7},
  pages={29639--29654},
  year={2019},
  publisher={IEEE}
}

@book{koenker2005quantile,
  title        = {Quantile Regression},
  author       = {Koenker, Roger},
  year         = {2005},
  series       = {Econometric Society Monographs},
  publisher    = {Cambridge University Press},
  address      = {Cambridge, UK},
  doi          = {10.1017/CBO9780511754098}
}

@article{shimura2012discretization,
  title={Discretization of distributions in the maximum domain of attraction},
  author={Shimura, Takaaki},
  journal={Extremes},
  volume={15},
  pages={299--317},
  year={2012},
  publisher={Springer}
}

@article{hitz2024extremes,
    author = {Adrien S. Hitz and Richard A. Davis and Gennady Samorodnitsky},
    title = {Discrete Extremes},
    journal = {Journal of Data Science},
    volume = {22},
    number = {4},
    year = {2024},
    pages = {524--536},
    doi = {10.6339/24-JDS1120},
    issn = {1680-743X},
    publisher = {School of Statistics, Renmin University of China}
}

@article{daouia2023extreme,
  title={Extreme value modelling of SARS-CoV-2 community transmission using discrete generalized {P}areto distributions},
  author={Daouia, Abdelaati and Stupfler, Gilles and Usseglio-Carleve, Antoine},
  journal={Royal Society Open Science},
  volume={10},
  number={3},
  pages={220977},
  year={2023},
  publisher={The Royal Society}
}

@article{ranjbar2022modelling,
  title={Modelling the extremes of seasonal viruses and hospital congestion: The example of flu in a Swiss hospital},
  author={Ranjbar, Setareh and Cantoni, Eva and Chavez-Demoulin, Val{\'e}rie and Marra, Giampiero and Radice, Rosalba and Jaton, Katia},
  journal={Journal of the Royal Statistical Society Series C: Applied Statistics},
  volume={71},
  number={4},
  pages={884--905},
  year={2022},
  publisher={Oxford University Press}
}

@article{prieto2014modelling,
  title={Modelling road accident blackspots data with the discrete generalized {P}areto distribution},
  author={Prieto, Faustino and G{\'o}mez-D{\'e}niz, Emilio and Sarabia, Jos{\'e} Mar{\'\i}a},
  journal={Accident Analysis \& Prevention},
  volume={71},
  pages={38--49},
  year={2014},
  publisher={Elsevier}
}

@article{wood2016smoothing,
  title={Smoothing parameter and model selection for general smooth models},
  author={Wood, Simon N and Pya, Natalya and S{\"a}fken, Benjamin},
  journal={Journal of the American Statistical Association},
  volume={111},
  number={516},
  pages={1548--1563},
  year={2016},
  publisher={Taylor \& Francis}
}

@article{ahmad2024extended,
  title={An extended generalized {P}areto regression model for count data},
  author={Ahmad, Touqeer and Gaetan, Carlo and Naveau, Philippe},
  journal={Statistical Modelling},
  pages={},
  year={2024},
  publisher={SAGE Publications Sage India: New Delhi, India}
}

@article{murphy2024automated,
  title={Automated threshold selection and associated inference uncertainty for univariate extremes},
  author={Murphy, Conor and Tawn, Jonathan A and Varty, Zak},
  journal={Technometrics},
  pages={1--10},
  year={2024},
  publisher={Taylor \& Francis}
}

@Manual{marra2024gjrm,
    title = {GJRM: Generalised Joint Regression Modelling (R package version 0.2-6.7)},
    year = {2024},
    author = {Giampiero Marra and Rosalba Radice},
  }

@article{watson2020weather,
    title = {{Weather-related power outage model with a growing domain: structure, performance, and generalisability}},
    year = {2020},
    journal = {The Journal of Engineering},
    author = {Watson, Peter L. and Cerrai, Diego and Koukoula, Marika and Wanik, David W. and Anagnostou, Emmanouil},
    number = {10},
    month = {10},
    pages = {817--826},
    volume = {2020},
    publisher = {The Institution of Engineering and Technology},
    url = {https://onlinelibrary.wiley.com/doi/full/10.1049/joe.2019.1274 https://onlinelibrary.wiley.com/doi/abs/10.1049/joe.2019.1274 https://ietresearch.onlinelibrary.wiley.com/doi/10.1049/joe.2019.1274},
    doi = {10.1049/JOE.2019.1274},
    issn = {2051-3305}
}

@article{alpay2009thunder,
    title = {{Dynamic modeling of power outages caused by thunderstorms}},
    year = {2020},
    journal = {Forecasting 2020, Vol. 2, Pages 151-162},
    author = {Alpay, Berk A. and Wanik, David and Watson, Peter and Cerrai, Diego and Liang, Guannan and Anagnostou, Emmanouil},
    number = {2},
    month = {5},
    pages = {151--162},
    volume = {2},
    publisher = {Multidisciplinary Digital Publishing Institute},
    url = {https://www.mdpi.com/2571-9394/2/2/8/htm https://www.mdpi.com/2571-9394/2/2/8},
    doi = {10.3390/FORECAST2020008},
    issn = {2571-9394}
}

@inproceedings{brester2020weather,
  title={Weather-based fault prediction in electricity networks with artificial neural networks},
  author={Brester, Christina and Niska, Harri and Ciszek, Robert and Kolehmainen, Mikko},
  booktitle={2020 IEEE Congress on Evolutionary Computation (CEC)},
  pages={1--8},
  year={2020},
  organization={IEEE}
}

@article{Tsioumpri2021WeatherNetworks,
    title = {{Weather related fault prediction in minimally monitored distribution networks}},
    year = {2021},
    journal = {Energies},
    author = {Tsioumpri, Eleni and Stephen, Bruce and McArthur, Stephen D. J.},
    number = {8},
    month = {4},
    pages = {2053},
    volume = {14},
    doi = {10.3390/en14082053},
    issn = {1996-1073}
}

@article{ghaemi2024stacking,
  title={A stacking-based fault forecasting study for power transmission lines under different weather conditions},
  author={Ghaemi, Ali and Safari, Amin and Quteishat, Anas and Younis, Mahmoud A},
  journal={Energy},
  volume={306},
  pages={132572},
  year={2024},
  publisher={Elsevier}
}

@article{Wilkinson2022ConsequenceStorms,
    title = {{Consequence forecasting: A rational framework for predicting the consequences of approaching storms}},
    year = {2022},
    journal = {Climate Risk Management},
    author = {Wilkinson, Sean and Dunn, Sarah and Adams, Russell and Kirchner-Bossi, Nicolas and Fowler, Hayley J. and Gonz{\'{a}}lez Ot{\'{a}}lora, Samuel and Pritchard, David and Mendes, Joana and Palin, Erika J. and Chan, Steven C.},
    pages = {100412},
    volume = {35},
    doi = {10.1016/j.crm.2022.100412},
    issn = {22120963}
}

@article{koenker1978regression,
  title={Regression quantiles},
  author={Koenker, Roger and Bassett Jr, Gilbert},
  journal={Econometrica: journal of the Econometric Society},
  pages={33--50},
  year={1978},
  publisher={JSTOR}
}

@article{zhu2024large,
  title={A Large-Scale Multi-Objective Evolutionary Quantile Estimation Model for Wind Power Probabilistic Forecasting},
  author={Zhu, Jianhua and He, Yaoyao},
  journal={IEEE Transactions on Evolutionary Computation},
  year={2024},
  publisher={IEEE}
}

@article{ruiz2024applications,
  title={Applications of Probabilistic Forecasting in Demand Response},
  author={Ruiz-Abell{\'o}n, Mar{\'\i}a Carmen and Fern{\'a}ndez-Jim{\'e}nez, Luis Alfredo and Guillam{\'o}n, Antonio and Gabald{\'o}n, Antonio},
  journal={Applied Sciences},
  volume={14},
  number={21},
  pages={9716},
  year={2024},
  publisher={MDPI}
}

@article{zhang2014review,
  title={Review on probabilistic forecasting of wind power generation},
  author={Zhang, Yao and Wang, Jianxue and Wang, Xifan},
  journal={Renewable and Sustainable Energy Reviews},
  volume={32},
  pages={255--270},
  year={2014},
  publisher={Elsevier}
}

@article{koenker2013distributional,
  title={Distributional vs. Quantile Regression},
  author={Koenker, R and Leorato, S and Peracchi, F and others},
  journal={CEIS RESEARCH PAPERS},
  volume={300},
  year={2013},
  publisher={Eief Working Papers Series}
}

@article{fasiolo2021fast,
  title={Fast calibrated additive quantile regression},
  author={Fasiolo, Matteo and Wood, Simon N and Zaffran, Margaux and Nedellec, Rapha{\"e}l and Goude, Yannig},
  journal={Journal of the American Statistical Association},
  volume={116},
  number={535},
  pages={1402--1412},
  year={2021},
  publisher={Taylor \& Francis}
}

@article{waldmann2013bayesian,
  title={Bayesian semiparametric additive quantile regression},
  author={Waldmann, Elisabeth and Kneib, Thomas and Yue, Yu Ryan and Lang, Stefan and Flexeder, Claudia},
  journal={Statistical Modelling},
  volume={13},
  number={3},
  pages={223--252},
  year={2013},
  publisher={SAGE Publications Sage India: New Delhi, India}
}

@book{dey2016extreme,
  title={Extreme value modeling and risk analysis: methods and applications},
  author={Dey, Dipak K and Yan, Jun},
  year={2016},
  publisher={CRC Press}
}

@article{machado2005quantiles,
  title={Quantiles for counts},
  author={Machado, Jos{\'e} A F and Silva, JMC Santos},
  journal={Journal of the American Statistical Association},
  volume={100},
  number={472},
  pages={1226--1237},
  year={2005},
  publisher={Taylor \& Francis}
}

@article{gonccalves2021forecasting,
  title={Forecasting conditional extreme quantiles for wind energy},
  author={Gon{\c{c}}alves, Carla and Cavalcante, Laura and Brito, Margarida and Bessa, Ricardo J and Gama, Joao},
  journal={Electric Power Systems Research},
  volume={190},
  pages={106636},
  year={2021},
  publisher={Elsevier}
}

@article{browell2021probabilistic,
  title={Probabilistic forecasting of regional net-load with conditional extremes and gridded {NWP}},
  author={Browell, Jethro and Fasiolo, Matteo},
  journal={IEEE Transactions on Smart Grid},
  volume={12},
  number={6},
  pages={5011--5019},
  year={2021},
  publisher={IEEE}
}

@article{fasiolo2021package,
 title={qgam: Bayesian Nonparametric Quantile Regression Modeling in {R}},
 volume={100},
 url={https://www.jstatsoft.org/index.php/jss/article/view/v100i09},
 doi={10.18637/jss.v100.i09},
 number={9},
 journal={Journal of Statistical Software},
 author={Fasiolo, Matteo and Wood, Simon N. and Zaffran, Margaux and Nedellec, Raphaël and Goude, Yannig},
 year={2021},
 pages={1–31}
}

@book{hastie1990gam,
  title={{G}eneralized {A}dditive {M}odels},
  author={Hastie, T. J. and Tibshirani, R. J.},
  year={1990},
  publisher={London:Chapman \& Hall}
}

@misc{Hersbach2018ERA5Present,
      title        = {{ERA5 hourly time-series data on single levels from 1940 to present}},
  year         = {2025},
  publisher    = {Copernicus Climate Change Service (C3S) Climate Data Store (CDS)},
  url          = {https://cds.climate.copernicus.eu/cdsapp#!/dataset/reanalysis-era5-single-levels},
  note         = {Accessed: 2025-04-07},
    author = {Hersbach, H. and Bell, B. and Berrisford, P. and Biavati, G. and Hor{\'{a}}nyi, A. and Mu{\~{n}}oz Sabater, J. and Nicolas, J. and Peubey, C. and Radu, R. and Rozum, I. and Schepers, D. and Simmons, A. and Soci, C. and Dee, D. and Th{\'{e}}paut, J-N.}
}

@article{manning2025antecedent,
  title={Antecedent rainfall, wind direction and seasonal effects may amplify the risk of wind-driven power outages in the UK},
  author={Manning, Colin and Wilkinson, Sean and Fowler, Hayley J and Kendon, Elizabeth J},
  journal={Communications Earth \& Environment},
  volume={6},
  number={1},
  pages={217},
  year={2025},
  publisher={Nature Publishing Group UK London}
}

@article{huang2025modeling,
  title={Modeling probabilistic micro-scale wind field for risk forecasts of power transmission systems during tropical cyclones},
  author={Huang, Xiubing and Wang, Naiyu},
  journal={Structural Safety},
  pages={102620},
  year={2025},
  publisher={Elsevier}
}

@article{souto2024identification,
  title={Identification of weather patterns and transitions likely to cause power outages in the United Kingdom},
  author={Souto, Laiz and Neal, Robert and Pope, James O and Gonzalez, Paula LM and Wilkinson, Jonathan and Taylor, Philip C},
  journal={Communications Earth \& Environment},
  volume={5},
  number={1},
  pages={49},
  year={2024},
  publisher={Nature Publishing Group UK London}
}

@article{ranjan2010combining,
  title={Combining probability forecasts},
  author={Ranjan, Roopesh and Gneiting, Tilmann},
  journal={Journal of the Royal Statistical Society Series B: Statistical Methodology},
  volume={72},
  number={1},
  pages={71--91},
  year={2010},
  publisher={Oxford University Press}
}

@article{raftery2005using,
  title={Using Bayesian model averaging to calibrate forecast ensembles},
  author={Raftery, Adrian E and Gneiting, Tilmann and Balabdaoui, Fadoua and Polakowski, Michael},
  journal={Monthly weather review},
  volume={133},
  number={5},
  pages={1155--1174},
  year={2005}
}

@book{coles_introduction_2001,
	title = {An introduction to statistical modeling of extreme values},
	publisher = {London: Springer},
	author = {Coles, Stuart},
	year = {2001},
}

@article{chavez2005generalized,
  title={Generalized additive modelling of sample extremes},
  author={Chavez-Demoulin, Val{\'e}rie and Davison, Anthony C},
  journal={Journal of the Royal Statistical Society Series C: Applied Statistics },
  volume={54},
  number={1},
  pages={207~--222},
  year={2005}
}

@article{davison_models_1990,
	title = {Models for {exceedances} over {high} {thresholds}},
	volume = {52},
	issn = {00359246},
	number = {3},
	urldate = {2021-08-16},
	journal = {Journal of the Royal Statistical Society. Series B (Methodological)},
	author = {Davison, A. C. and Smith, R. L.},
	year = {1990},
	pages = {393--442},
}

@inproceedings{leutbecher2009diagnosis,
  title={Diagnosis of ensemble forecasting systems},
  author={Leutbecher, Martin},
  booktitle={Seminar on diagnosis of forecasting and data assimilation systems},
  pages={235--266},
  year={2009}
}

@article{gneiting2007strictly,
  title={Strictly proper scoring rules, prediction, and estimation},
  author={Gneiting, Tilmann and Raftery, Adrian E},
  journal={Journal of the American statistical Association},
  volume={102},
  number={477},
  pages={359--378},
  year={2007},
  publisher={Taylor \& Francis}
}

@article{jantre2023bayesian,
  title={Bayesian quantile regression for longitudinal count data},
  author={Jantre, Sanket},
  journal={Journal of Statistical Computation and Simulation},
  volume={93},
  number={1},
  pages={103--127},
  year={2023},
  publisher={Taylor \& Francis}
}

@article{cha2025assessing,
  title={Assessing the cascading impacts of natural hazards on Critical National Infrastructure ({CNI}) using Scotland as a case study},
  author={Cha, Younghwa and White, Christopher J and Gonzalez, Paula LM and Wallace, Emily and Harkin, David and Brett, Lou and Mattu, Kanzis and Smith, R Sean and Thom, Craig and Alix, Medlyn D and others},
  journal={npj Natural Hazards},
  volume={2},
  number={1},
  pages={108},
  year={2025},
  publisher={Nature Publishing Group UK London}
}

@article{gneiting2011comparing,
  title={Comparing density forecasts using threshold-and quantile-weighted scoring rules},
  author={Gneiting, Tilmann and Ranjan, Roopesh},
  journal={Journal of Business \& Economic Statistics},
  volume={29},
  number={3},
  pages={411--422},
  year={2011},
  publisher={Taylor \& Francis}
}

@article{allen2024weighted,
  title={Weighted scoringrules: Emphasizing particular outcomes when evaluating probabilistic forecasts},
  author={Allen, Sam},
  journal={Journal of Statistical Software},
  volume={110},
  pages={1--26},
  year={2024}
}

\clearpage
\setcounter{subsection}{0}
\renewcommand\thesubsection{\Alph{subsection}}
\renewcommand\thefigure{\thesubsection.\arabic{figure}}
\renewcommand\thetable{\thesubsection.\arabic{table}}

\section*{Appendices}\label{sec:appendix}

This appendix presents additional simulation experiments and detailed model specifications employed in the real data application.  The first Appendix~\ref{app:notation_metrics} describe the notation used across the manuscript to easier the reference. The Appendix~\ref{app:sim_three_scenario} part examines the robustness of the \textit{X-flexForecast} method in scenarios where the weather-related covariates are misspecified. Despite the model misspecification, \textit{X-flexForecast} continues to yield coherent forecasts of the predictive distribution. The Appendix~\ref{app:sensibility_threshold} presents further details on how \textit{X-flexForecast} behaves under different choices of the probability level $\alpha^{(T)}$ associated with the threshold between the bulk quantile–regression component and the tail. Lastly, Appendix~\ref{app:flex_model_specifications} outlines the cross-validated procedure used to select the transition threshold $\alpha^{(T)}_{d,t}$ and the covariate sets DGP components of \textit{X-flexForecast} models presented in Section~\ref{sec::results}, jointly with additional metrics for the model evaluation as the macro-micro AUC and the threshold weighted continuous ranked probability score.

\subsection{Notation and Evaluation Metrics}
\setcounter{figure}{0}
\setcounter{table}{0}
\label{app:notation_metrics}

This appendix consolidates the notation introduced in Section~\ref{sec:background} and the definitions of the evaluation metrics used in Section~\ref{sec:simulation} and Section~\ref{sec::application}. Indices refer to district $d$ and time $t$ unless otherwise stated. Bold symbols denote vectors; calligraphic letters denote sets or collections. For notation from cited bibliography, it is recommended to verify the original works


\setlength{\LTpre}{6pt}
\setlength{\LTpost}{6pt}
\renewcommand{\arraystretch}{1.05}

\begin{longtable}{p{3.8cm}p{11.0cm}}
\caption{Notation used throughout the paper.}\label{tab:notation_long}\\
\toprule
\textbf{Symbol} & \textbf{Description} \\
\midrule
\endfirsthead
\toprule
\textbf{Symbol} & \textbf{Description} \\
\midrule
\endhead
\midrule
\multicolumn{2}{r}{\textit{(continued on next page)}} \\
\bottomrule
\endfoot
\bottomrule
\endlastfoot

\multicolumn{2}{l}{\textit{Indices, sets, and data}}\\
$ d \in \{1,\ldots,D\} $ & District index. \\
$ t \in \{1,\ldots,T\} $ & Time index (day or lead-time stamp). \\
$ \mathcal{D} $ & Training dataset across all districts and times. \\

\multicolumn{2}{l}{\textit{Responses and covariates}}\\
$ Y_{d,t} \in \{0,1,2,\ldots\} $ & Random variable representing the fault count at $(d,t)$. \\
$ y_{d,t} \in \{0,1,2,\ldots\} $ & Observed fault count at $(d,t)$. \\
$ \mathbf{x}_{d,t} $ & Covariate vector at $(d,t)$ (weather features). \\
$ \mathbf{x}^{S_k}_{d,t} $ & Subset of covariates $\mathbf{x}_{d,t}$ retrieved from $S_{k}$ set of indexes. \\
$ \mathbf{A}_{d,t}$ &  is a column vector containing the subset of covariates \\

\multicolumn{2}{l}{\textit{Predictive distribution and quantiles}}\\
$ F(Y_{d,t}\mid \mathbf{x}_{d,t}) $ & Conditional CDF of $Y_{d,t}$ given $\mathbf{x}_{d,t}$. \\
$ q_{\alpha}(Y_{d,t}\mid \mathbf{x}_{d,t}) $ & Conditional $\alpha$-quantile at level $\alpha\in [0,1)$. \\
$ \widehat{F}_{QR} $ & CDF induced by the fitted additive quantile regression (bulk). \\

\multicolumn{2}{l}{\textit{Additive quantile regression (bulk)}}\\
$ \beta_{\alpha,0} $ & Intercept at quantile level $\alpha$. \\
$ \bm{\beta}_{\alpha,A} $ & Coefficients for $\mathbf{A}_{d,t}$ at level $\alpha$. \\
$ f_{\alpha,k}(\mathbf{x}^{S_k}_{d,t}) $ & Smooth component $k$ at level $\alpha$. \\
$ b_{k,j}(\cdot),\; K_k $ & Basis function $j=1,\dots,K_k$ for smooth $k$. \\
$ \gamma_{\alpha,k,j} $ & Coefficient of basis $b_{k,j}$ at level $\alpha$. \\
$ g^{-1}(\cdot) $ & Inverse link used in the additive predictor. \\
$ \lambda>0 $ & Smoothing parameter for the smoothed pinball loss. \\
$ \boldsymbol{\theta}_{B} $ & Parameter set for the bulk (QR/GAM) component. \\[2pt]

\multicolumn{2}{l}{\textit{Tail modeling: Generalized Pareto (GP) and Discrete Generalized Pareto (DGP)}}\\

$ u_{d,t} $ & Threshold to calculate exceedances value at $(d,t)$. \\
$ \alpha^{(T)}_{d,t} $ & Mixture/threshold quantile level . \\
$ \phi\in(0,1) $ & Target exceedance probability. \\
$ \sigma>0,\; \xi $ & Scale and shape parameters from tail models, respectively. \\
$\mathbf{x}^{T_{k}}_{d,t}$ & Subset of weather predictors used only $\sigma(\mathbf{x}^{T}_{d,t})$ indicated by the $T_{k}$ indexes. \\
$ F_{\text{GP}}(\cdot) $ & Continuous GP CDF. \\
$ F_{\text{DGP}}(\cdot) $ & DGP CDF on integer support. \\
$ \lfloor \cdot \rfloor,\; \lceil \cdot \rceil $ & Floor and ceiling operators. \\
$ \boldsymbol{\theta}_T $ & Parameter set for the tail (DGP) component. \\[2pt]

\multicolumn{2}{l}{\textit{X-flexForecast}}\\
$ \widehat{F}_{d,t} $ & Predictive CDF from \textit{X-flexForecast} at $(d,t)$. \\
$ \tilde{\alpha}^{(T)}_{d,t}(\mathbf{x}_{d,t}) $ & CDF mass of $\widehat{F}_{QR}$ corresponding to $\lfloor q_{\alpha^{(T)}_{d,t}}(\mathbf{x}_{d,t}) \rfloor$. \\
$ \text{HRES} $ & Deterministic high-resolution forecast. \\
$ \hat{f}_j $ & Estimate obtained from the fitted model $j$. \\[2pt]
$ \tau^{(AG)}_{d,t}; \tau^{(RA)}_{d,t} $ & Fault-count thresholds defining Green/Amber and Red/Amber bands, respectively. \\

\multicolumn{2}{l}{\textit{Quantile and loss notation}}\\
$ \rho_{\alpha}(u) $ & Pinball loss: $u\big(\alpha-\mathbb{I}\{u<0\}\big)$. \\
\\
\multicolumn{2}{l}{\textit{Examples of weather covariates (elements of $\mathbf{x}_{d,t}$)}}\\
$ t2m $ & 2-meter air temperature (K). \\
$ tp $ & Total precipitation (mm) \\
$ ws10_{\max} $ & Max 10-meter wind speed (m/s). \\
\small{$cape\_mean\_ge100\_covar$} & Average CAPE exceedance of $\ge 100$ J/kg. \\
$ tp_{q_{90}} $ & 90th percentile of precipitation across ensemble members. \\

\end{longtable}


\begin{longtable}{p{4.0cm}p{10.8cm}}
\caption{Metrics used in simulations and the case study.}\label{tab:metrics_long}\\
\toprule
\textbf{Metric} & \textbf{Definition / Use} \\
\midrule
\endfirsthead
\toprule
\textbf{Metric} & \textbf{Definition / Use} \\
\midrule
\endhead
\midrule
\multicolumn{2}{r}{\textit{(continued on next page)}} \\
\bottomrule
\endfoot
\bottomrule
\endlastfoot

Pinball (Quantile) Loss
& For residual $u=y-q_{\alpha}$, $\rho_{\alpha}(u)=u\big(\alpha-\mathbb{I}\{u<0\}\big)$.
Used to train and assess quantile forecasts for the bulk model. \\[0.35em]

Smoothed Pinball Loss
& Differentiable variant used  with smoothing parameter $\lambda>0$; reduces to standard pinball as $\lambda\to 0$ (see \cite{fasiolo2021fast}). \\[0.35em]

RMSE
& Root mean squared error for point predictions (e.g., BMA mean):
$\text{RMSE}=\sqrt{n^{-1}\sum_{i=1}^n(\hat{y}_i-y_i)^2}$. \\[0.35em]

MAE
& Mean absolute error: $\text{MAE}=n^{-1}\sum_{i=1}^n |\hat{y}_i-y_i|$. \\[0.35em]

Brier Score (BS)
& For a binary event $Z$, $\text{BS}=(\hat{p}-\mathbb{I}\{Z\})^2$.
Applied to categorical bands via one-vs-rest events (e.g., Amber). \\[0.35em]

AUC
& Area under the ROC curve for classification of band-specific events (e.g., Amber vs.\ not Amber), using predicted event probabilities from $\widehat{F}_{d,t}$. \\[0.35em]

Reliability calibration
& Empirical coverage of nominal prediction intervals derived from $\widehat{F}_{d,t}$; used in reliability summaries. \\[0.35em]

twCRPS
& Threshold-weighted continuous ranked probability score:
$\mathrm{twCRPS}(F,y)=\int_{0}^{\infty} w(z)\big(F(z)-\mathbb{I}\{y\le z\}\big)^2\,\mathrm{d}z$,
with weight function $w(z)=\mathbb{I}\{z\ge u\}$.
Emphasizes forecast performance in the upper tail; evaluated from Monte Carlo samples of $\widehat{F}_{d,t}$ using the \texttt{scoringExtra} implementation. \\[0.35em]

\end{longtable}

\newpage

\subsection{Third scenario: misspecified model}
\setcounter{figure}{0}
\setcounter{table}{0}
\label{app:sim_three_scenario}

Given that the true relationship between conditional quantiles and available predictors is typically unknown, it is essential to evaluate the impact of model misspecification. To this end, the predictors $\mathbf{Z} = (\mathbf{z}^{\{1\}}, \mathbf{z}^{\{2\}})$ are sampled following the setup in Section~\ref{sec:sim_two}, while the response variable $y$ is generated using the simulation design from Section~\ref{sec:sim_one}, ensuring independence from $\mathbf{z}^{\{2\}}$. This induces a misspecification of the model by incorporating a predictor that does not influence the distribution of $y$. Specifically, a basis expansion of dimension $k=10$ is calculated over $\mathbf{z}^{\{2\}}$ and included in the parameterization of the conditional quantiles and $\log{\left(\sigma(\mathbf{z})\right)}$ for both \textit{X-FlexForecast} and EDGP, although $\mathbf{z}^{\{2\}}$ is irrelevant to the true data generation process. Table \ref{tab:sim_scenario_three} summarizes the results, showing that RMSE-based comparisons remain consistent with previous findings: \textit{X-FlexForecast} and competing methods exhibit similar precision for bulk quantiles, while \textit{X-FlexForecast} continues to achieve superior performance in the estimation of tail quantiles.

\begin{table}[ht]
    \vspace{-0.75cm}
    \centering
    \resizebox{\textwidth}{!}{
    \begin{tabular}{c|cccc|cccc}
        \multicolumn{9}{c}{} \\
        \multicolumn{1}{c}{} & \multicolumn{4}{c}{$\xi = 0.0$} & \multicolumn{4}{c}{$\xi = 0.3$} \\
        \hline
        $\alpha$ & $\bar{q}(\mathbf{X})$ & \textit{X-flexForecast} & \textit{qgam} & EDGP & $\bar{q}(\mathbf{X})$ & \textit{X-flexForecast} & \textit{qgam} & EDGP \\
        \hline
        &  & \multicolumn{3}{c|}{RMSE} &  & \multicolumn{3}{c}{RMSE} \\
        \hline
        0.05   & 0.09  & \textbf{0.61 $\pm$ 0.41} & --- & 0.52 $\pm$ 0.06 & 0.09 & 0.60 $\pm$ 0.35 & --- & \textbf{0.53 $\pm$ 0.07} \\
        0.25   & 1.39  & \textbf{1.23 $\pm$ 0.42} & --- & 0.96 $\pm$ 0.05 & 1.41 & 1.22 $\pm$ 0.33 & --- & \textbf{0.96 $\pm$ 0.14} \\
        0.50   & 3.94  & 1.53 $\pm$ 0.52 & --- & \textbf{1.03 $\pm$ 0.11} & 3.97 & 1.51 $\pm$ 0.46 & --- & \textbf{1.04 $\pm$ 0.34} \\
        0.75   & 7.96  & 1.95 $\pm$ 0.65 & --- & \textbf{1.21 $\pm$ 0.29} & 8.00 & 1.91 $\pm$ 0.63 & --- & \textbf{1.24 $\pm$ 0.68} \\
        0.95   & 14.42 & \textbf{1.98 $\pm$ 1.02} & 2.26 $\pm$ 0.64 & 3.24 $\pm$ 0.77 & 14.93 & \textbf{2.10 $\pm$ 0.99} & 2.32 $\pm$ 0.65 & 4.19 $\pm$ 1.65 \\
        0.99   & 18.45 & \textbf{2.01 $\pm$ 1.02} & 4.00 $\pm$ 0.84 & 10.97 $\pm$ 2.08 & 21.48 & \textbf{2.56 $\pm$ 0.99} & 4.01 $\pm$ 0.89 & 13.11 $\pm$ 3.46 \\
        0.999  & 24.17 & \textbf{2.60 $\pm$ 1.33} & 17.23 $\pm$ 5.30 & 22.25 $\pm$ 4.09 & 38.79 & \textbf{5.01 $\pm$ 2.51} & 13.07 $\pm$ 8.32 & 26.17 $\pm$ 6.56 \\
        0.9999 & 30.03 & \textbf{3.85 $\pm$ 2.38} & 69.74 $\pm$ 17.36 & 33.47 $\pm$ 6.13 & 73.28 & \textbf{14.46 $\pm$ 9.53} & 42.62 $\pm$ 18.77 & 44.55 $\pm$ 8.13 \\
        \hline
    \end{tabular}}
    \caption{Average RMSE over 1000 replications for the third simulation scenario. Bold values indicate the lowest RMSE among the three approaches. Columns labeled as $\bar{q}(\mathbf{X})$ show the average true quantile at level $\alpha$. Values for bulk quantiles are omitted for \textit{qgam}, as they are identical to those of \textit{X-flexForecast}.}
    \label{tab:sim_scenario_three}
\end{table}

\subsection{Sensitivity Analysis of Threshold Selection in \textit{X-flexForecast}}
\setcounter{figure}{0}
\setcounter{table}{0}
\label{app:sensibility_threshold}
\addcontentsline{toc}{section}{Appendix new}

This appendix provides additional details on the behaviour of \textit{X-flexForecast} with respect to the choice of the probability level $\alpha^{(T)}$ that determines the transition between the bulk quantile–regression component and the discrete generalized Pareto (DGP) tail. In practical forecasting settings, the true exceedance probability is unknown, and the operative threshold must be selected. The analysis below examines the stability of the predictive distribution and the corresponding tail–parameter estimates under deliberate choices of $\alpha^{(T)}$.

\subsubsection{Estimated versus data-generating thresholds}

Although Section~\ref{sec:simulation} specifies $\alpha^{(T)} = 1-\phi$ to align the model with the simulation design, the practical integer threshold is
\[
    \hat{u}(\bm{x}_{d,t}) = 
    \left\lceil 
        \hat{q}_{\alpha^{(T)}_{d,t}}(\bm{x}_{d,t})
    \right\rceil,
\]
and therefore differs from the data-generating threshold
\[
    u
    = \inf\{ k\in\mathbb{N}_0 :
        F_{\mathrm{dg}}(k \mid \boldsymbol{\theta}_B )
        \ge 1-\phi
      \}.
\]
Two structural features contribute to this discrepancy. First, $\hat{q}_{\alpha^{(T)}}(\bm{x}_{d,t})$ is subject to estimation errors from the quantile–regression fit. Second, the distribution is discrete: several nearby probability levels may correspond to the same integer quantile, and hence to the same operative threshold. As a result, $\alpha^{(T)}$ could be interpreted as determining a neighborhood of admissible thresholds rather than a single probability level.

\subsubsection{Grid of candidate thresholds}\label{app:appendix_grid_threshold}

To assess sensitivity to threshold misspecification, we extend all the settings from the simulation of Section~\ref{sec:sim_one} with $\phi=0.1$ by considering a grid of candidate transition probabilities
\[
\mathcal{A}=\{(1-\phi)-0.05,(1-\phi)-0.04,\dots,(1-\phi)+0.04,(1-\phi)+0.05\},
\]
with increments of $0.01$. The step size of 0.01 and the use of ten candidate values define a sensible local grid for $\alpha^{(T)}$, reflecting a standard hyperparameter-tuning trade-off that provides sufficient exploration of the parameter space while avoiding an unnecessarily dense search, with the range and spacing adaptable to the scope of the analysis and desired granularity. For each $\alpha^{(T)}\in\mathcal{A}$, the corresponding threshold $\hat{u}$ is recomputed from the estimated quantile, the tail model is refitted using the implied exceedances, and predictive accuracy is evaluated via the RMSE of the estimated tail quantiles $\hat{q}_\alpha(\bm{x})$ relative to the true quantiles for $\alpha\in\{0.95,0.99,0.999,0.9999\}$. Figure~\ref{fig:threshold_0_9_scaled_rmse_tail_quantiles} summarizes the resulting scaled RMSE patterns, where the RMSE is scaled by the average of the corresponding true quantile values. For intermediate upper quantiles ($\alpha=0.95,0.99$), RMSE varies only mildly across $\mathcal{A}$, reflecting that multiple probability levels often induce the same integer threshold and that exceedances remain within a range where the GP approximation is adequate. In contrast, for far-tail quantiles ($\alpha=0.999,0.9999$), RMSE increases sharply when $\alpha^{(T)}$ is set too high, due to the reduced number of exceedances and the resulting increase in variance of $(\hat{\sigma},\hat{\xi})$. Slightly lower values of $\alpha^{(T)}$ can partially offset this effect by increasing the exceedance sample size, whereas overly low thresholds introduce substantial bias by contaminating tail estimation with non-extreme observations.

\begin{figure}[htb]
    \centering
    \includegraphics[width=\linewidth]{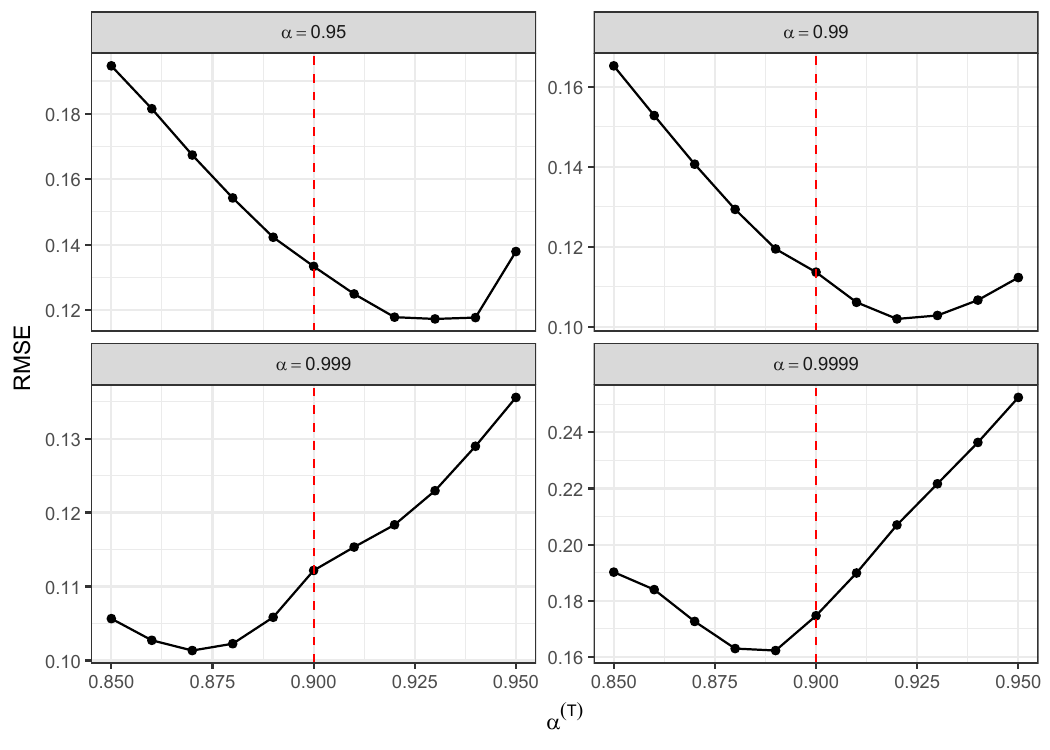}
    \caption{Scaled Average RMSE of the predicted tail quantiles for $\alpha\in{0.95,0.99,0.999,0.9999}$ across the candidate threshold levels $\alpha^{(T)}\in\mathcal{A}$, under the simulation design with $\phi=0.1$ -- indicated by the dashed line. 
    }
    \label{fig:threshold_0_9_scaled_rmse_tail_quantiles}
\end{figure}

\subsubsection{Tail-parameter estimates}\label{app:appendix_tail_parameters_estimates}

To complement the RMSE results, we examine the mean and standard deviation of the estimated tail parameters $(\hat{\xi},\hat{\sigma})$ for each $\alpha^{(T)}\in\mathcal{A}$. The empirical means in Figure~\ref{fig:mean_parameters} indicate that $\hat{\xi}$ remains stable for thresholds corresponding to probability levels at or above approximately $0.89$, while the mean of $\hat{\sigma}$ remains close to the value associated with the data-generating threshold. Figure~\ref{fig:sd_parameters} shows an increase in the standard deviation of both parameters as $\alpha^{(T)}$ increases, consistent with the reduction in the number of observations within the exceedances set. The dashed (blue) line marks the estimates obtained when the true exceedance threshold $u$ is used, while the dash–dot (orange) line indicates the true parameter value

\begin{figure}[htpb]
    \centering
    \includegraphics[width=\linewidth]{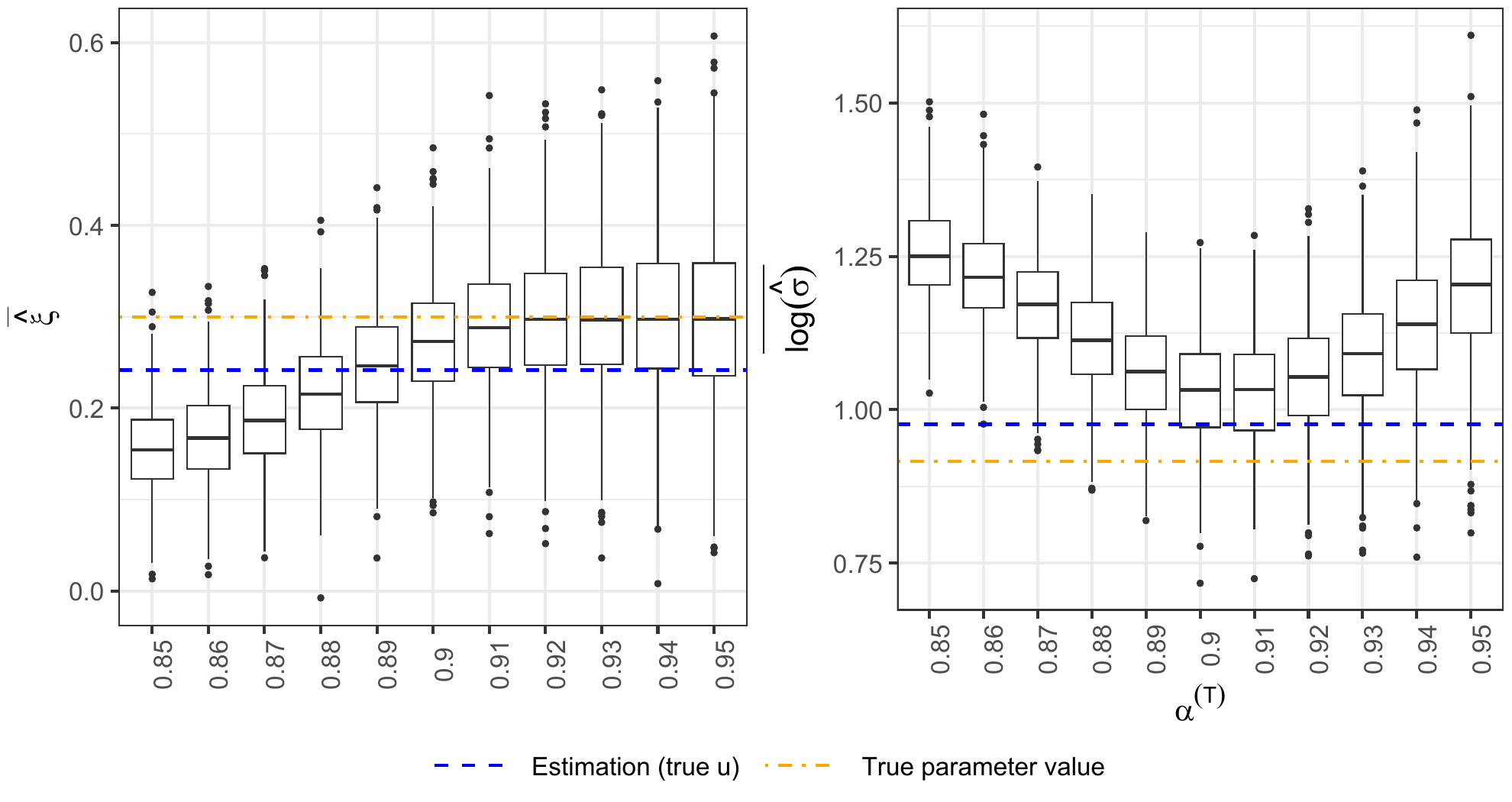}
    \caption{Mean estimates of the shape parameter $\hat{\xi}$ and log--scale parameter $\hat{\sigma}$ across candidate thresholds $\alpha^{(T)}$ obtained from the tail model of \textit{X-flexForecast} derived from simulations described in Appendix~\ref{app:sensibility_threshold} with $\phi=0.1$. 
    }
    \label{fig:mean_parameters}
\end{figure}

\begin{figure}[htbp]
    \centering
    \includegraphics[width=\linewidth]{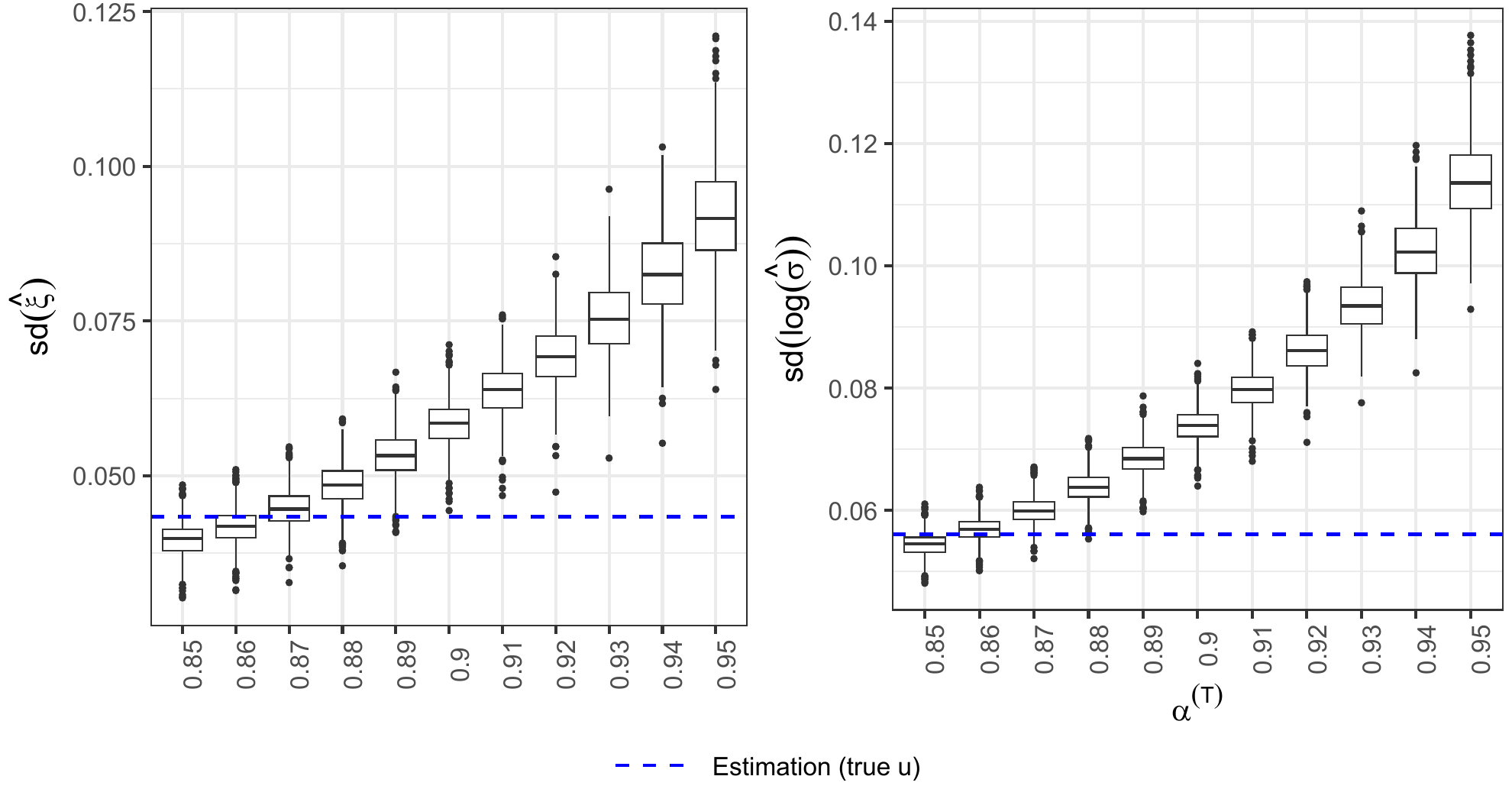}
    \caption{Standard deviation of the estimated shape parameter $\hat{\xi}$ and log--scale parameter $\hat{\sigma}$ across candidate thresholds $\alpha^{(T)}$ from simulations described in Appendix~\ref{app:sensibility_threshold} with $\phi = 0.1$.  
    }
    \label{fig:sd_parameters}
\end{figure}




Across all experiments, the predictive performance of \textit{X-flexForecast} is stable for thresholds within a neighbourhood around the data-generating value. Thresholds set substantially below this region introduce bias by incorporating non-extreme observations into the tail model, while thresholds set substantially above it reduce the number of exceedances and increase estimator variance.  
Within the practically relevant range, the method provides reliable estimates for both bulk and tail quantiles, in agreement with classical threshold-selection principles for exceedance models.

\subsubsection{Additional exceedance probabilities}\label{sec:appendix_extra_exceedance_prob}

To assess generality, the analysis was replicated for $\phi \in \{0.15,0.20,0.25\}$. 
The resulting patterns mirror those observed for $\phi = 0.1$---see 
Sections~\ref{app:appendix_grid_threshold} and~\ref{app:appendix_tail_parameters_estimates} for details---exhibiting stable behaviour for intermediate thresholds, variance inflation for overly high $\alpha^{(T)}$, and bias for overly low $\alpha^{(T)}$.  
These results support the robustness of the approach across a range of plausible choices for $\alpha^{(T)}$.

\begin{figure}[htb]
    \centering
    \includegraphics[width=\linewidth]{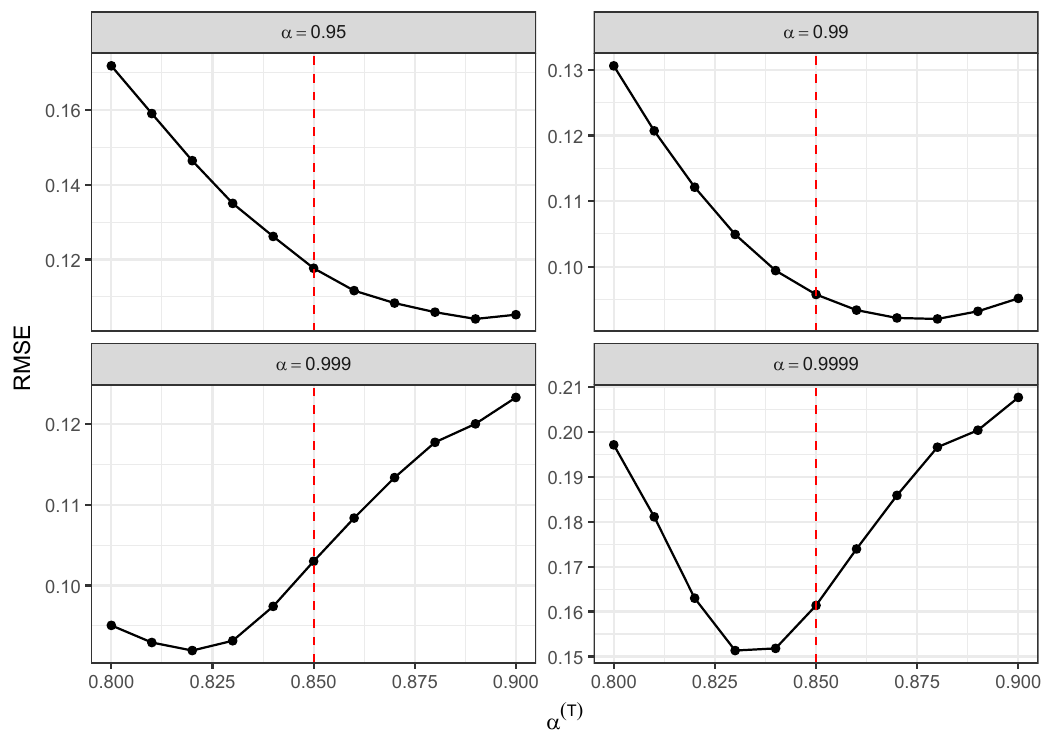}
    \caption{Scaled average RMSE of the predicted tail quantiles for $\alpha\in{0.95,0.99,0.999,0.9999}$ across the candidate threshold levels $\alpha^{(T)}\in\mathcal{A}$, under the simulation design with $\phi=0.15$ -- indicated by the dashed line. 
    }
    \label{fig:threshold_0_85_scaled_rmse_tail_quantiles}
\end{figure}

\begin{figure}[htpb]
    \centering
    \includegraphics[width=\linewidth]{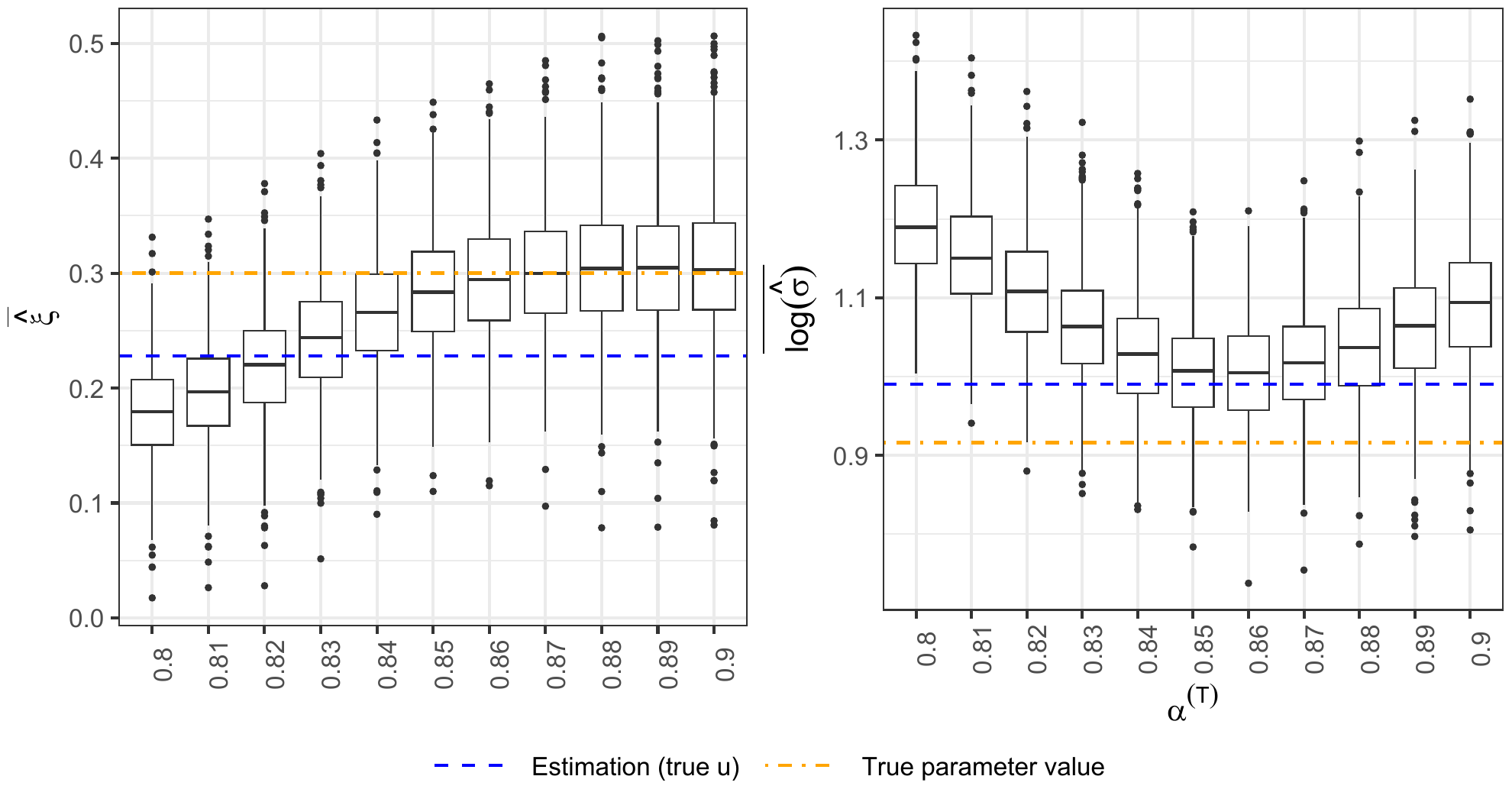}
    \caption{Mean estimates of the shape parameter $\hat{\xi}$ and log--scale parameter $\hat{\sigma}$ across candidate thresholds $\alpha^{(T)}$ obtained from the tail model of \textit{X-flexForecast} derived from simulations described in Appendix~\ref{app:sensibility_threshold} with $\phi=0.15$. 
    }
    \label{fig:mean_parameters_0.85}
\end{figure}

\begin{figure}[htbp]
    \centering
    \includegraphics[width=\linewidth]{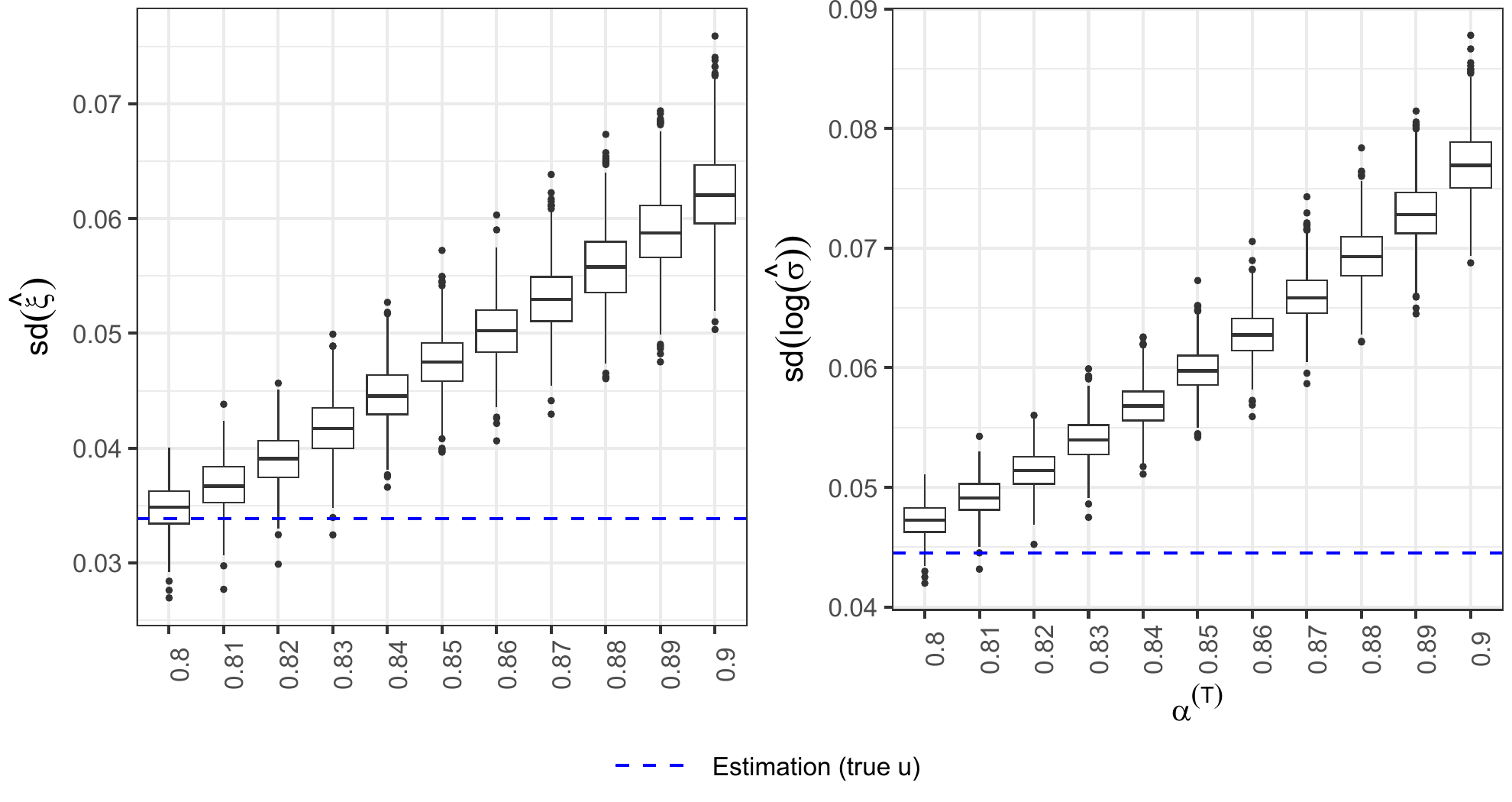}
    \caption{Standard deviation of the estimated shape parameter $\hat{\xi}$ and log--scale parameter $\hat{\sigma}$ across candidate thresholds $\alpha^{(T)}$ from simulations described in Appendix~\ref{app:sensibility_threshold} with $\phi = 0.15$.  
    }
    \label{fig:sd_parameters_0.85}
\end{figure}

\begin{figure}[htb]
    \centering
    \includegraphics[width=\linewidth]{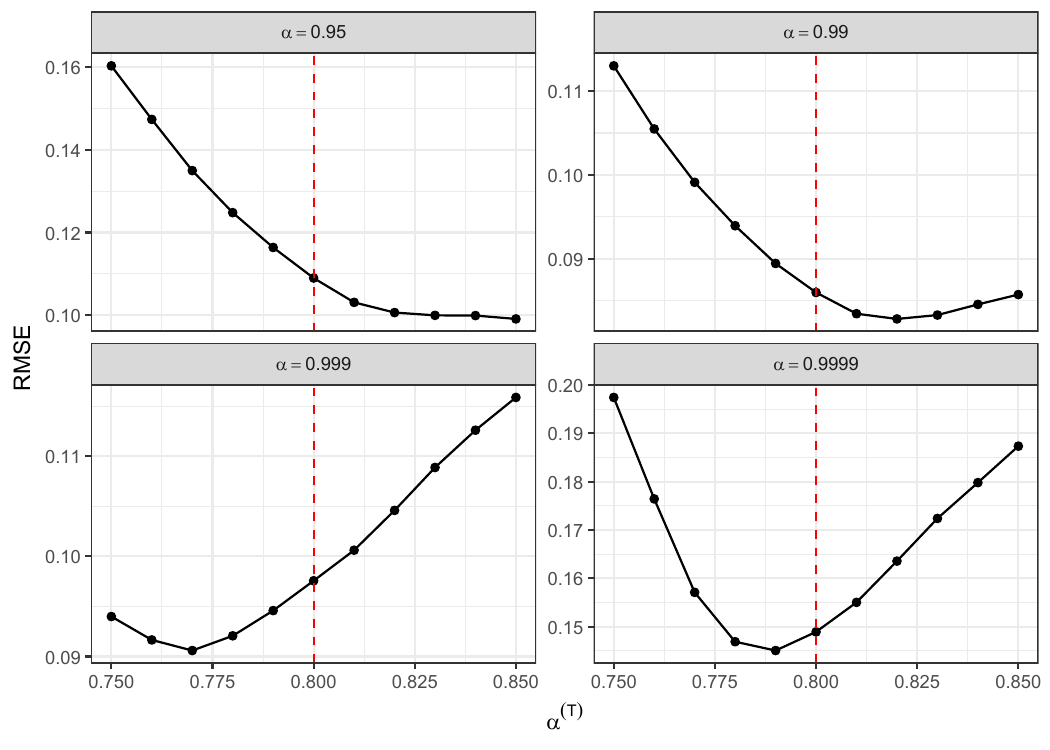}
    \caption{Scaled average RMSE of the predicted tail quantiles for $\alpha\in{0.95,0.99,0.999,0.9999}$ across the candidate threshold levels $\alpha^{(T)}\in\mathcal{A}$, under the simulation design with $\phi=0.2$ -- indicated by the dashed line. 
    }
    \label{fig:threshold_0_8_scaled_rmse_tail_quantiles}
\end{figure}

\begin{figure}[htpb]
    \centering
    \includegraphics[width=\linewidth]{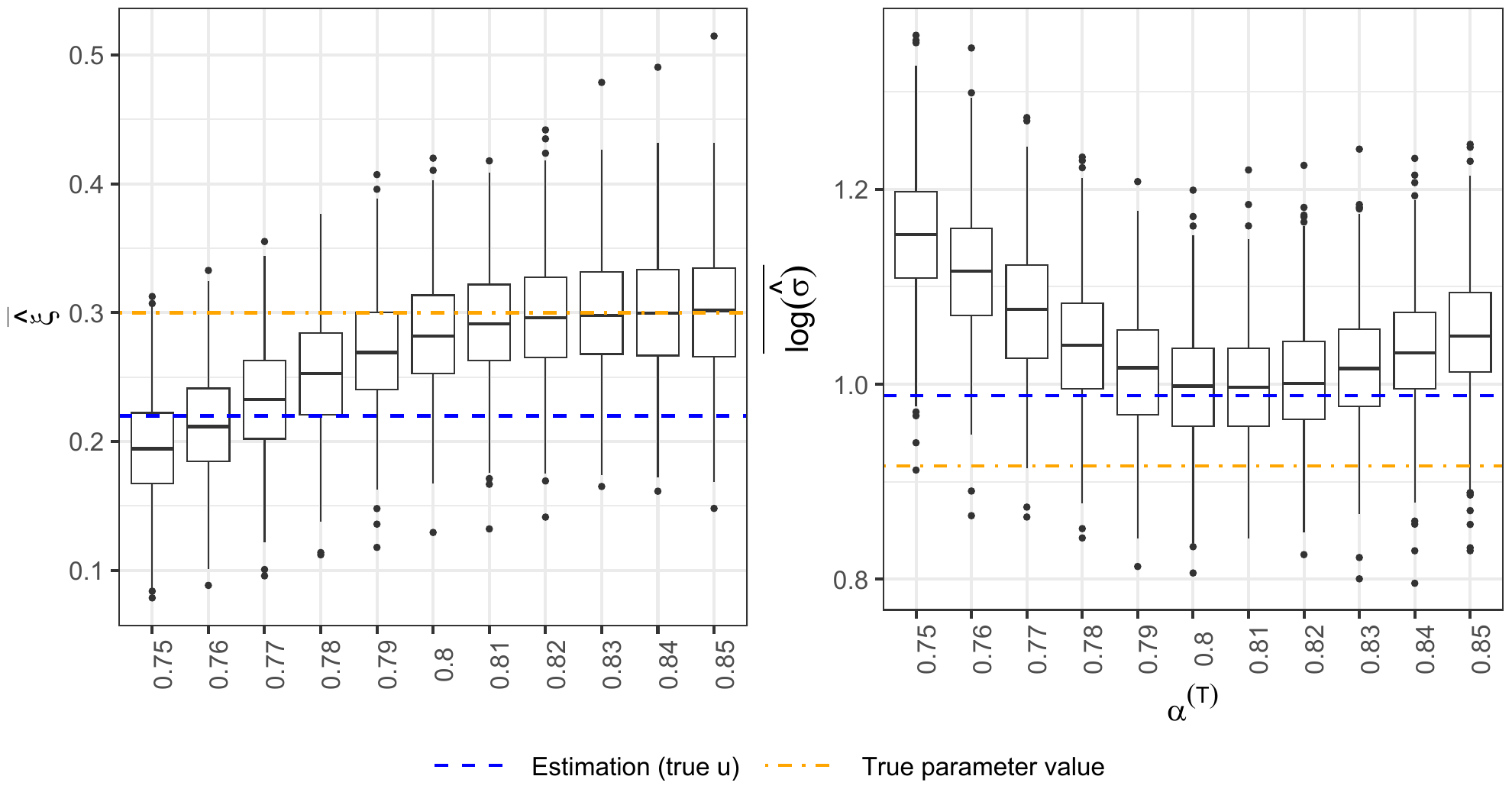}
    \caption{Mean estimates of the shape parameter $\hat{\xi}$ and log--scale parameter $\hat{\sigma}$ across candidate thresholds $\alpha^{(T)}$ obtained from the tail model of \textit{X-flexForecast} derived from simulations described in Appendix~\ref{app:sensibility_threshold} with $\phi=0.2$. 
    }
    \label{fig:mean_parameters_0.8}
\end{figure}

\begin{figure}[htbp]
    \centering
    \includegraphics[width=\linewidth]{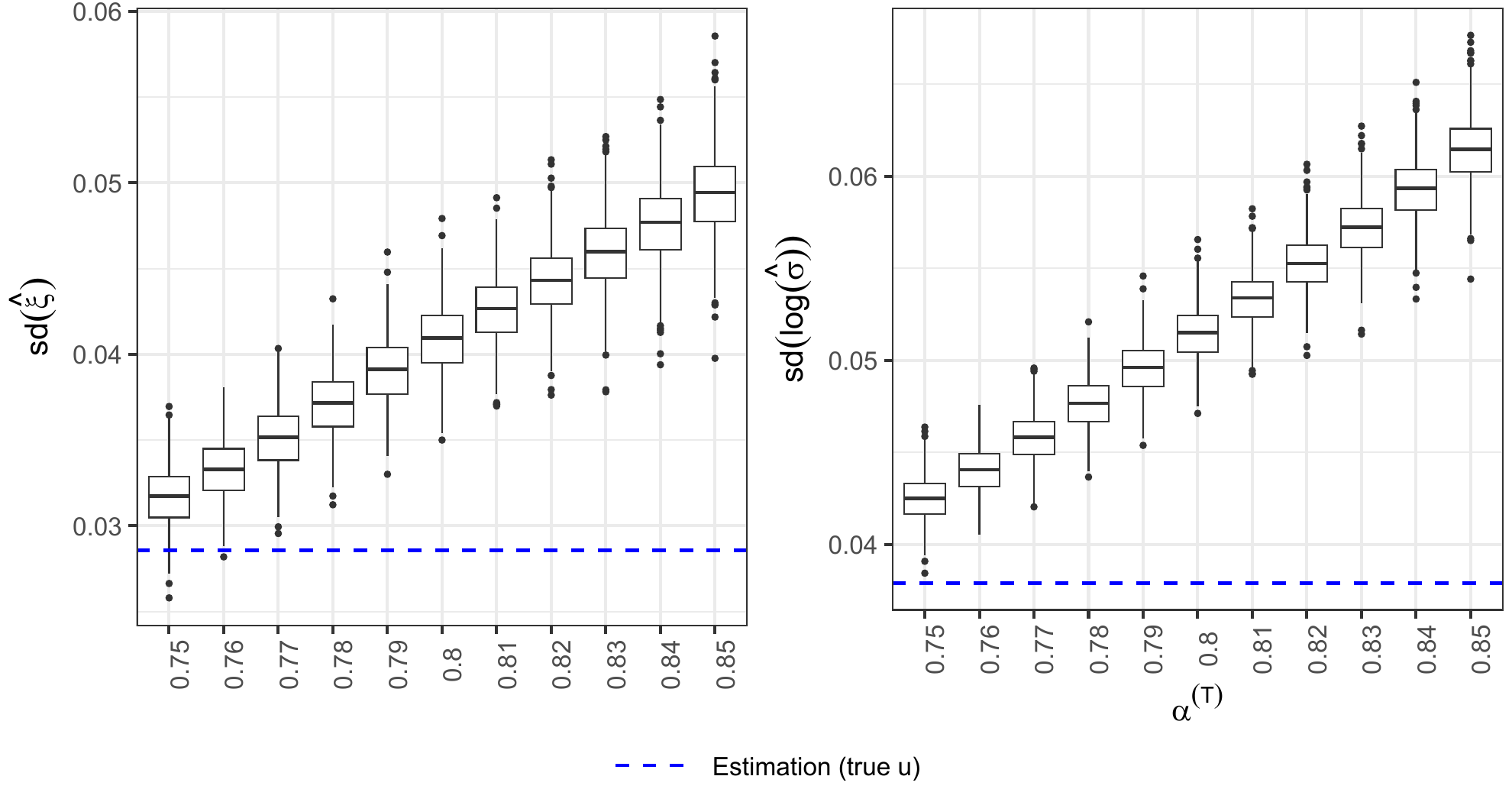}
    \caption{Standard deviation of the estimated shape parameter $\hat{\xi}$ and log--scale parameter $\hat{\sigma}$ across candidate thresholds $\alpha^{(T)}$ from simulations described in Appendix~\ref{app:sensibility_threshold} with $\phi = 0.2$.  
    }
    \label{fig:sd_parameters_0.8}
\end{figure}

\begin{figure}[htb]
    \centering
    \includegraphics[width=\linewidth]{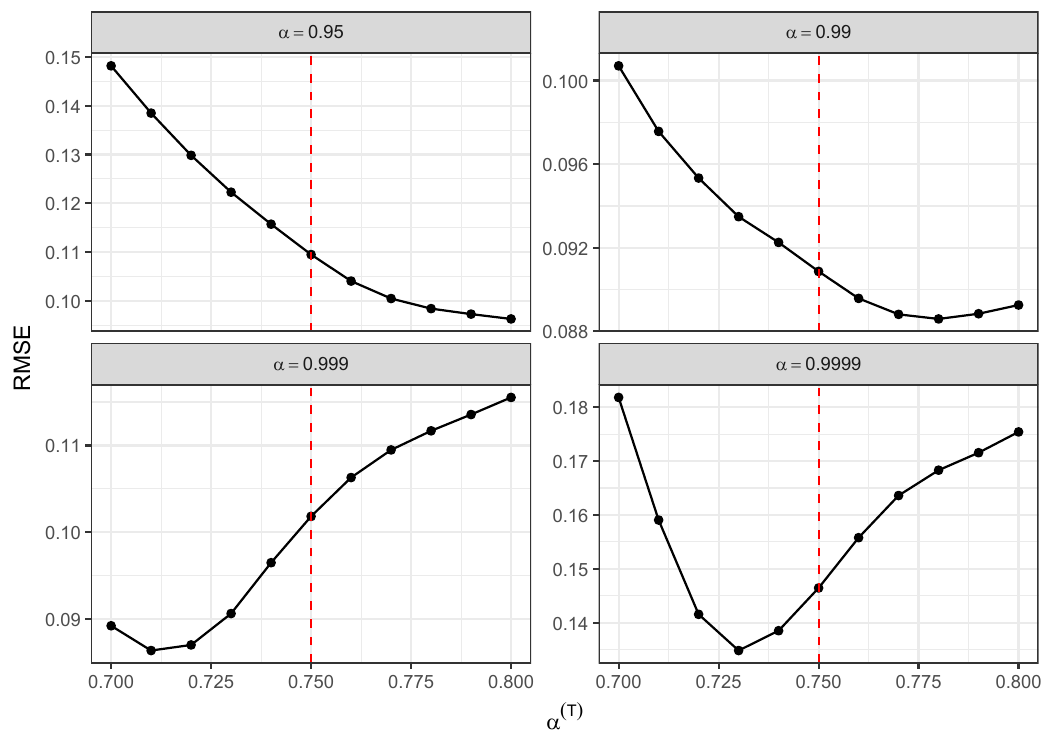}
    \caption{Scaled average RMSE of the predicted tail quantiles for $\alpha\in{0.95,0.99,0.999,0.9999}$ across the candidate threshold levels $\alpha^{(T)}\in\mathcal{A}$, under the simulation design with $\phi=0.25$ -- indicated by the dashed line. 
    }
    \label{fig:threshold_0_75_scaled_rmse_tail_quantiles}
\end{figure}

\begin{figure}[htpb]
    \centering
    \includegraphics[width=\linewidth]{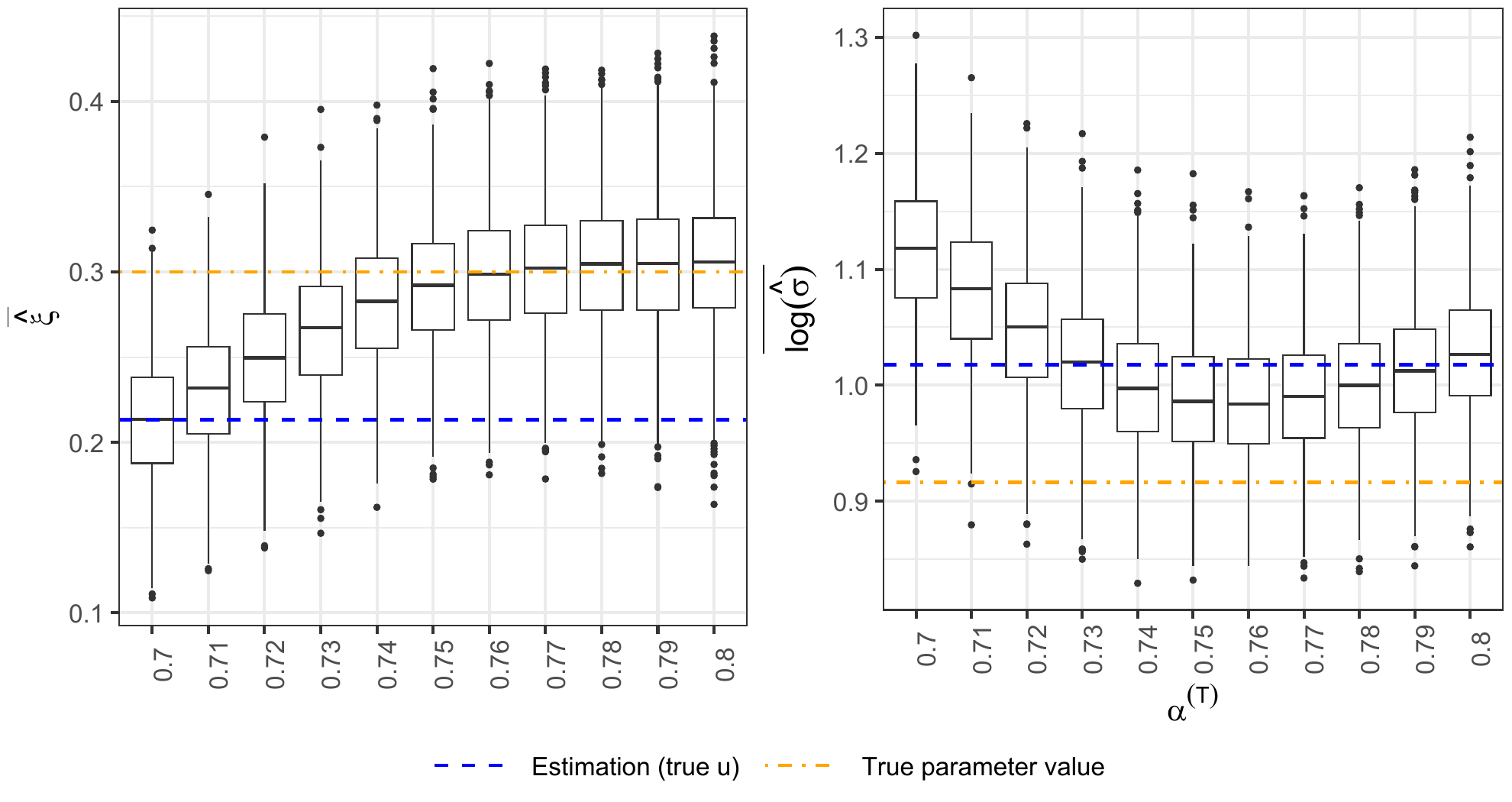}
    \caption{Mean estimates of the shape parameter $\hat{\xi}$ and log--scale parameter $\hat{\sigma}$ across candidate thresholds $\alpha^{(T)}$ obtained from the tail model of \textit{X-flexForecast} derived from simulations described in Appendix~\ref{app:sensibility_threshold} with $\phi=0.25$. 
    }
    \label{fig:mean_parameters_0.75}
\end{figure}

\begin{figure}[htbp]
    \centering
    \includegraphics[width=\linewidth]{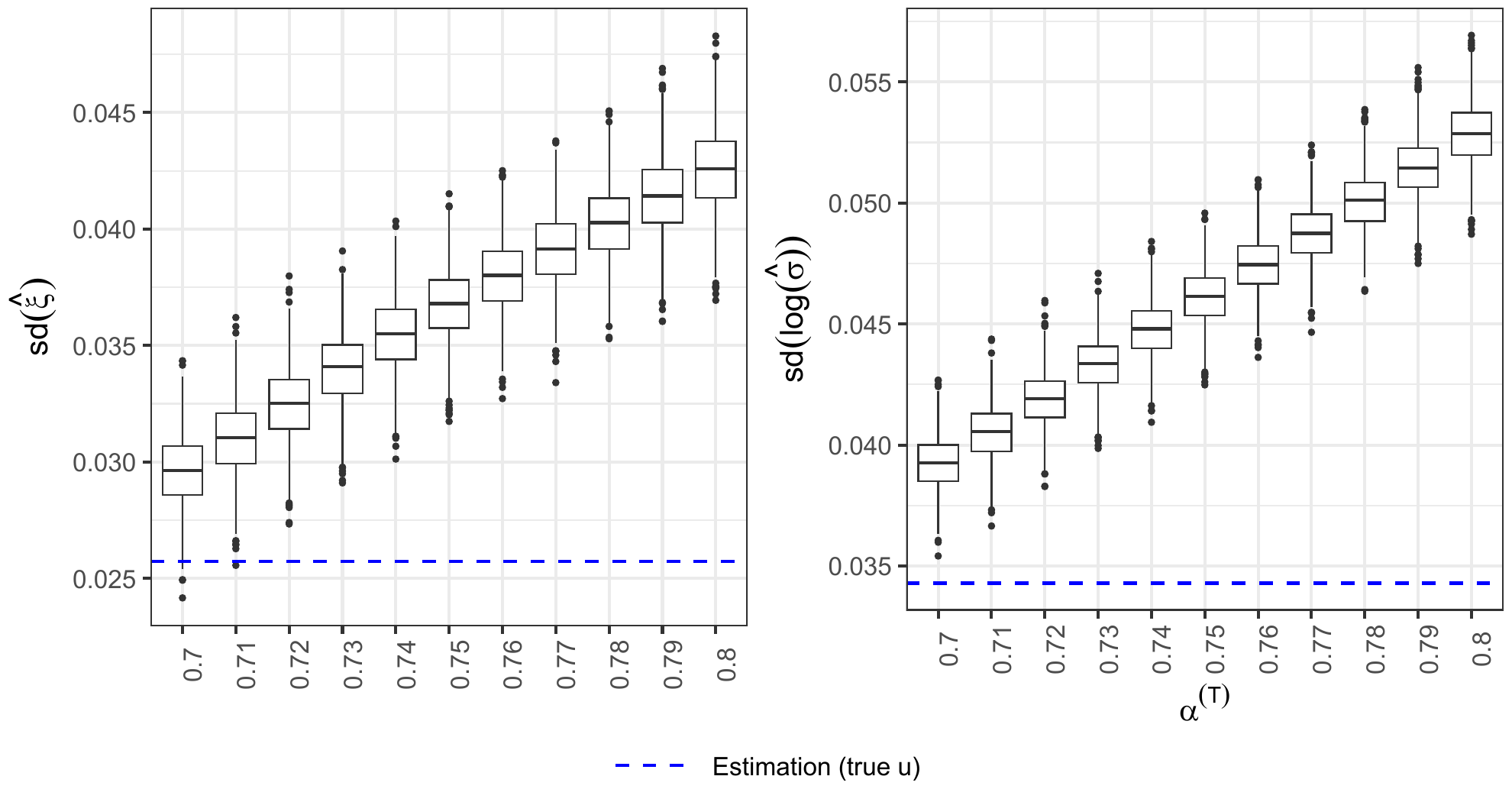}
    \caption{Standard deviation of the estimated shape parameter $\hat{\xi}$ and log--scale parameter $\hat{\sigma}$ across candidate thresholds $\alpha^{(T)}$ from simulations described in Appendix~\ref{app:sensibility_threshold} with $\phi = 0.25$.  
    }
    \label{fig:sd_parameters_0.75}
\end{figure}

\clearpage
\subsection{\textit{X-flexForecast} model specifications}
\setcounter{figure}{0}
\setcounter{table}{0}
\label{app:flex_model_specifications}

\addcontentsline{toc}{section}{Appendix}

The \textit{X-flexForecast} model, described in Section~\ref{sec:background} and summarized in Equation~\eqref{eq:mixture}, requires specification of both the transition probability $\alpha^{(T)}_{d,t}$ defining the bulk–tail mixture and the set of covariates $\mathbf{x}^{(DGP)}_{d,t}$ used to model the scale parameter $\sigma(\mathbf{x}^{T}_{d,t})$ of the discrete generalized Pareto (DGP) distribution, for each district $d$ and time resolution $t$. Using the dataset introduced in Section~\ref{sec::application}, these quantities were selected via a 14-fold cross-validation scheme, with each fold corresponding to a full regulatory year (April–March), thereby ensuring that each fold contains at least one complete winter season.

Candidate specifications were compared based on out-of-sample predictive performance rather than through optimization of a differentiable loss function, as the transition probability $\alpha^{(T)}_{d,t}$ cannot be expressed as a continuous decision variable in a formal learning objective. Instead, $\alpha^{(T)}_{d,t}$ was treated as a hyperparameter and selected from a discrete grid of candidates. For each candidate threshold, predictive skill was evaluated using cross-validated scoring rules, including the Brier score (BS) and the area under the ROC curve (AUC), computed separately for the Green, Amber, and Red categories. The selected threshold $\widehat{\alpha}^{(T)}_{d,t}$ corresponds to the configuration achieving the largest average improvement in BS and AUC across categories; when the two criteria disagreed, BS was prioritized as the primary selection metric, as it is a strictly proper scoring rule assessing calibration and accuracy, whereas AUC primarily reflects ranking performance \citep{gneiting2007strictly}. This criteria is also used to define the subset $\mathbf{x}_{d,t}^{T}$. The final combination of weather predictors, mixture threshold $\alpha^{(T)}_{d,t}$, and covariates for the DGP component selected under this procedure is summarized in Table~\ref{tab:selected_hyper}.

\begin{table}[htb]
\centering
\footnotesize
\caption{DGP model specifications for each district $d$ at time resolution $t=24$.}
\begin{tabular}{l p{6cm} c}
\toprule
$d$ & $\mathbf{x}^{(DGP)}_{d,t}$ & $\alpha^{(T)}_{d,t}$ \\
\midrule
Ayrshire \& Clyde South & \gls{ws10max} $+$ $tp_{q_{90}}$ & 0.85 \\
Central \& Fife & \gls{ws10max} & 0.80 \\
Dee Valley \& Mid Wales & \gls{ws10max} $+$ $tp_{q_{90}}$ & 0.90 \\
Dumfries \& Stranraer & \gls{ws10max} $+$ $tp_{q_{90}}$ & 0.90 \\
Edinburgh \& Borders & \gls{ws10max} & 0.90 \\
Lanarkshire & $tp_{q_{90}}$ & 0.90 \\
Merseyside & $t2m_{\text{mean}}$ & 0.75 \\
Mid Cheshire & $tp_{q_{90}} + \text{cape\_mean}_{\ge100}$ & 0.85 \\
North Wales & \gls{ws10max} $+$ $tp_{q_{90}}$ & 0.90 \\
Wirral & \gls{ws10max} $+$ $t2m_{\text{mean}}$ & 0.80 \\
\bottomrule
\end{tabular}
\label{tab:selected_hyper}
\end{table}

AUC values are nevertheless reported as complementary performance measures, as they facilitate interpretation by operational stakeholders familiar with this metric (see Section~\ref{sec::application} and Appendix~\ref{app:flex_model_specifications}). Table~\ref{tab:extra_auc_table} reports micro- and macro-averaged AUC values across districts and methods.

\begin{table}[htb]
\centering
\small
\caption{Micro- and macro-averaged AUC by district for the \textit{mqgam} model and the \textit{X-flexForecast} model.}
\begin{tabular}{lccccc}
\toprule
& \multicolumn{2}{c}{\textit{qgam}} & \multicolumn{2}{c}{\textit{X-flexForecast}} \\
\cmidrule(lr){2-3} \cmidrule(lr){4-5}
District & AUC (macro) & AUC (micro) & AUC (macro) & AUC (micro) \\
\midrule
Ayrshire \& Clyde South     & 0.858 & 0.996 & 0.867 & 0.996 \\
Central \& Fife             & 0.766 & 0.993 & 0.812 & 0.995 \\
Dee Valley \& Mid Wales     & 0.844 & 0.992 & 0.854 & 0.993 \\
Dumfries \& Stranraer       & 0.857 & 0.993 & 0.883 & 0.993 \\
Edinburgh \& Borders        & 0.790 & 0.989 & 0.829 & 0.990 \\
\addlinespace
Lanarkshire                 & 0.755 & 0.993 & 0.767 & 0.993 \\
Merseyside                  & 0.637 & 0.986 & 0.704 & 0.988 \\
Mid Cheshire                & 0.839 & 0.992 & 0.821 & 0.992 \\
North Wales                 & 0.837 & 0.990 & 0.869 & 0.990 \\
Wirral                      & 0.728 & 0.988 & 0.739 & 0.989 \\
\bottomrule
\end{tabular}
\label{tab:extra_auc_table}
\end{table}

In addition to the BS for assessment of probabilistic forecasting, we also evaluate the models using other proper scoring rules. Following the reviewer’s suggestion, we compute the continuous ranked probability score (CRPS), and in particular its threshold-weighted variant (twCRPS; \citealp{gneiting2011comparing}), implemented via the \texttt{scoringRules} package \citep{allen2024weighted}. This choice reflects the structure of our mixture model: on the region of the support where $y \le \lfloor q_{\alpha^{(T)}_{d,t}}(\mathbf{x}_{t}) \rfloor$, the forecasts from the multiple additive quantile regression model and from \textit{X-flexForecast} coincide. An unweighted CRPS would therefore place substantial emphasis on parts of the predictive distribution where the two models are identical. To focus the comparison on the upper tail—where the forecasting approaches differ—we use a twCRPS with weight function $w(z) = \mathbb{I}\{ z \ge a \}$, where $a = \lfloor q_{\alpha^{(T)}_{d,t}}(\mathbf{x}_{t}) \rfloor$ and $b=\infty$, so that the evaluation isolates differences in tail behavior. Because the predictive distributions $\widehat{F}_{d,t}(\mathbf{x}_{d,t})$ do have a closed-form expressions the twCRPS is computed via \texttt{twcrps\_sample}, which estimates the score from Monte Carlo samples drawn from $\widehat{F}_{d,t}(\mathbf{x}_{d,t})$ under each model. We use $n_{\text{samples}} = 100{,}000$ samples for numerical stability. Further methodological details are provided in \cite{allen2024weighted}. Figure~\ref{fig:twcrps_ss} shows the empirical density of the twCRPS skill scores. The distribution places most of its mass to the right of the zero reference line, indicating that, for the majority of cases, the \textit{X-flexForecast} model achieves lower twCRPS values—and therefore higher predictive skill—than the standard multiple \textit{qgam} approach. We note that, for a small number of districts and on fewer than four days in total, an instability was observed in the twCRPS-based comparison. In these cases, the threshold used in the twCRPS lay above the effective support of the \textit{qgam} predictive distribution. Because the \textit{qgam} model can exhibit relatively short upper tails, its empirical CDF reached one early within the weighted region, causing the twCRPS to collapse to zero and the resulting skill score to become artificially large or unstable. This behavior reflects a structural truncation of the \textit{qgam} tail rather than a genuine improvement in predictive performance. These instances were omitted from the corresponding plots for visual clarity, but are fully accounted for in the underlying analysis. Given their rarity and limited impact, this phenomenon does not affect the overall conclusions and is reported here for completeness.

\begin{figure}
    \centering
    \includegraphics[width=\linewidth]{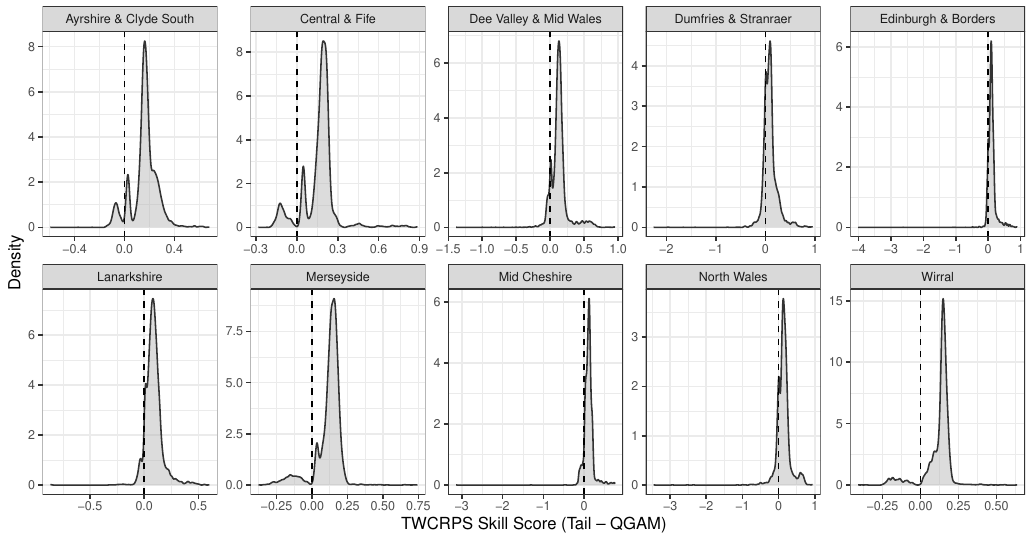}
    \caption{Empirical densities of the twCRPS skill scores using the \textit{qgam} as reference model across the ten districts. The vertical dashed line marks zero, corresponding to equal performance between the two approaches. In every district, the distribution places the majority of its mass on the positive side, indicating that the \textit{X\textnormal{-}flexForecast} tail model attains lower (i.e., better) threshold-weighted CRPS values for most forecast instances relative to the standard multiple \textit{qgam} model.}
    \label{fig:twcrps_ss}
\end{figure}


\end{document}